\newif\iflongversion
\longversiontrue
\iflongversion
\documentclass[sigconf, natbib=false, balance=false]{acmart}
\else
\documentclass[sigconf, review, natbib=false, balance=false]{acmart}
\fi
\acmConference{SPLC}{October 2020}{Montreal, Canada}
\acmYear{2020}

\usepackage{mysty-red}

\newcommand*\uli[1]{}

\begin{document}

\title{Featured Games}

\author{Uli Fahrenberg}%
\affiliation{%
  \institution{{\'E}cole polytechnique, France}}%
\email{uli@lix.polytechnique.fr}%

\author{Axel Legay}
\affiliation{%
  \institution{Universit{\'e} Catholique de Louvain, Belgium}}%
\email{axel.legay@uclouvain.be}

\begin{abstract}
  Feature-based SPL analysis and family-based model checking have seen
  rapid development.  Many model checking problems can be reduced to
  two-player games on finite graphs.  A prominent example is
  mu-calculus model checking, which is generally done by translating
  to parity games, but also many quantitative model-checking problems
  can be reduced to (quantitative) games.

  In their FASE'20 paper, ter Beek et al.\ introduce parity games with
  variability in order to develop family-based mu-calculus model
  checking of featured transition systems.  We generalize their model
  to general featured games and show how these may be analysed in a
  family-based manner.

  We introduce featured reachability games, featured minimum
  reachability games, featured discounted games, featured energy
  games, and featured parity games.  We show how to compute winners
  and values of such games in a family-based manner.  We also show
  that all these featured games admit optimal featured strategies,
  which project to optimal strategies for any product.  Further, we
  develop family-based algorithms, using late splitting, to compute
  winners, values, and optimal strategies for all the featured games
  we have introduced.
\end{abstract}



\maketitle

\section{Introduction}

Managing variability between products is a key challenge in software
product line (SPL) engineering.  In feature-based SPL analysis,
products are abstracted into features, so that any product is a
combination of a set of given features, specifying characteristics
that are present or absent in the particular product.

\emph{Featured transition systems} (FTS), introduced by Classen
\etal~\cite{DBLP:conf/icse/ClassenHSLR10}, are high-level
representations of SPL which allow for model checking of qualitative
and quantitative properties of SPL.  Model checking is an established
technique for verifying the behavior of complex systems, and SPL model
checking is an active research
subject~\cite{DBLP:conf/icse/ClassenHSLR10,
  DBLP:journals/sttt/ClassenCHLS12, DBLP:conf/splc/CordyCHSL13,
  DBLP:conf/splc/BeekM14, DBLP:journals/fac/ChrszonDKB18,
  DBLP:journals/tse/BeekLLV20, DBLP:journals/scp/ClassenCHLS14,
  DBLP:journals/jlp/LochauMBR16, DBLP:conf/fase/BeekVW17}.

The number of products in an SPL grows exponentially with the number
of features, hence model checking each individual product is
prohibitive.  Thus, \emph{family-based model checking} has been
introduced in~\cite{DBLP:conf/icse/ClassenHSLR10}, allowing for the
simultaneous verification of all products.  The family-based approach
has seen rapid development~\cite{DBLP:conf/icse/ApelRWGB13,
  DBLP:conf/fase/BeekVW17, DBLP:journals/sttt/ClassenCHLS12,
  DBLP:journals/scp/ClassenCHLS14, DBLP:journals/tse/ClassenCSHLR13,
  DBLP:conf/icse/ClassenHSL11, DBLP:conf/fase/BeekLVW20} and has been
extended to conformance model
checking~\cite{DBLP:conf/icse/CordyCPSHL12}, abstraction-based model
checking~\cite{DBLP:conf/sigsoft/CordyHLSDL14,
  DBLP:conf/fase/DimovskiLW19, DBLP:conf/fase/DimovskiW17}, real-time
formalisms~\cite{DBLP:journals/jlp/BeoharM16,
  DBLP:conf/splc/CordySHL12}, probabilistic
systems~\cite{DBLP:conf/splc/BeekLLV15,
  DBLP:journals/fac/ChrszonDKB18, DBLP:conf/sigsoft/CordyHLSDL14,
  DBLP:conf/hase/RodriguesANLCSS15}, and quantitative model
checking~\cite{DBLP:conf/splc/OlaecheaFAL16,
  DBLP:journals/sttt/FahrenbergL19};
see~\cite{DBLP:conf/birthday/CordyDLPCHSR19} for a recent
survey.

Many model checking problems can be reduced to two-player \emph{games}
on finite graphs.  A prominent example is $\mu$-calculus model
checking, which is generally done by translating to parity
games~\cite{DBLP:reference/mc/BradfieldW18}, but also many
quantitative model-checking problems can be reduced to (quantitative)
games, see~\cite{DBLP:conf/ictac/FahrenbergLQ19,
  DBLP:journals/tcs/FahrenbergL14}.

\begin{figure}[tbp]
  \centering
  \begin{tikzpicture}[x=2cm]
    \node[state, initial] (0) at (0,0) {$s_0$};
    \node[state] (1) at (1,0) {$s_1$};
    \node[state] (2) at (2,0) {$s_2$};
    \path (0) edge node[above] {$\ins\mid \ltrue$} (1);
    \path (1) edge node[above] {$\ins\mid \dollar$} (2);
    \path (1) edge[out=120, in=60] node[above] {$\std\mid \euro$} (0);
    \path (2) edge[out=210, in=330] node[below] {$\xxl\mid \ltrue$}
    (0);
  \end{tikzpicture}
  \caption{FTS model $S$ of a simple coffee machine SPL.}
  \Description{FTS model $S$ of a simple coffee machine SPL.}
  \label{fi:beek-fts}
\end{figure}
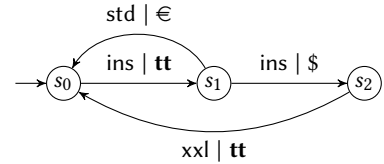

In their recent paper~\cite{DBLP:conf/fase/BeekLVW20}, ter~Beek \etal
introduce a procedure for family-based $\mu$-calculus model checking
of FTS.  They define a translation to \emph{parity games with
  variability} and then develop an algorithm for family-based analysis
of such games.  We give an example inspired
by~\cite{DBLP:conf/fase/BeekLVW20}.  Figure~\ref{fi:beek-fts} shows a
toy model $S$ of a coffee machine with feature set
$\{ \euro, \dollar\}$ and three products $\{ \euro\}$, $\{ \dollar\}$,
and $\{ \euro, \dollar\}$.  The machine can accept coins at the $\ins$
transitions, deliver regular coffee at the $\std$ transition, and hand
out extra large coffee at the $\xxl$ transition; but the $\std$
transition is only enabled if the $\euro$ feature is present, and the
second $\ins$ transition exists only if the $\dollar$ feature is
present.

\begin{figure}[tbp]
  \centering%
  $\phi= \nu X. \mu Y. \overbrace{%
    \big( \underbrace{%
      (\langle \ins\rangle Y\lor\langle \xxl\rangle Y)}_{ \phi_3}%
    \lor\langle \std\rangle X \big)}^{ \phi_2}$
  \caption{$\mu$-calculus specification for $S$.}
  \Description{$\mu$-calculus specification for $S$.}
  \label{fi:beek-mu}
\end{figure}

In Fig.~\ref{fi:beek-mu} we define a $\mu$-calculus formula $\phi$
which expresses the property that there exists an infinite run of the
system along which infinitely many regular coffees are delivered.  We
quickly recall the translation introduced
in~\cite{DBLP:conf/fase/BeekLVW20}, which is a feature-enriched
version of the standard
translation~\cite{DBLP:reference/mc/BradfieldW18} from $\mu$-calculus
model checking to parity games.

Let $N$ be a set of features, $\Sigma$ a set of actions, and
$F=( S, i, T, \gamma)$ an FTS, with states $S$, initial state
$i\in S$, transitions $T\subseteq S\times \Sigma\times S$, and feature
guards $\gamma: T\to \Bool( N)$, the set of boolean expressions over
$N$.  Let $\phi$ be a $\mu$-calculus formula and denote by
$\sub{ \phi}$ the set of subformulas of $\phi$ (including $\phi$
itself).  The featured parity game associated with $F$ and $\phi$ has
states $S\times \sub{ \phi}$, with initial state $( i, \phi)$, and the
owners, priorities and successors of states are given in
Fig.~\ref{fi:beek-trans}, where $\ad{ X}{ \psi}$ denotes the
alternation depth of variable $X$ in formula $\psi$.

\begin{figure}[t]
  \centering
  \Description{Translation to featured parity games
    from~\cite{DBLP:conf/fase/BeekLVW20}.}
  \caption{Translation to featured parity games
    from~\cite{DBLP:conf/fase/BeekLVW20}.}
  \begin{tabular}{l|c|l|l}
    State & Owner & Successors & Priority \\\hline
    $( s, \lfalse)$ & 1 & $\emptyset$ & 0 \\
    $( s, \ltrue)$ & 2 & $\emptyset$ & 0 \\
    $( s, \psi_1\lor \psi_2)$ & 1 & $\{( s, \psi_1)_{ / \ltrue}, ( s,
    \psi_2)_{ / \ltrue}\}$ & 0 \\
    $( s, \psi_1\land \psi_2)$ & 2 & $\{( s, \psi_1)_{ / \ltrue}, ( s,
    \psi_2)_{ / \ltrue}\}$ & 0 \\
    $( s, \langle a\rangle \psi)$ & 1 & $\{( s', \psi)_{ /
      \gamma}\mid s\ttto{ \gamma}{ a} s'\}$ & 0 \\
    $( s, [ a] \psi)$ & 2 & $\{( s', \psi)_{ / \gamma}\mid s\ttto{
      \gamma}{ a} s'\}$ & 0 \\
    $( s, \nu X. \psi)$ & 2 & $\{( s, \psi[ X:= \nu x. \psi])_{ /
      \ltrue}\}$ & $2\lfloor \ad{ X}{ \psi}/2 \rfloor$ \\
    $( s, \mu X. \psi)$ & 2 & $\{( s, \psi[ X:= \mu x. \psi])_{ /
      \ltrue}\}$ & $2\lfloor \ad{ X}{ \psi}/2 \rfloor+ 1$
  \end{tabular}
  \label{fi:beek-trans}
\end{figure}

\begin{figure}[b]
  \centering
  \begin{tikzpicture}[x=1.4cm, y=1.4cm]
    \node[state, diamond] (00) at (0,0) {$0$};
    \node[state, diamond] (10) at (1,0) {$0$};
    \node[state, diamond] (40) at (4,0) {$0$};
    \node[state, diamond] (01) at (0,1) {$0$};
    \node[state, diamond] (11) at (1,1) {$0$};
    \node[state, rectangle] (21) at (2,1) {$1$};
    \node[state, diamond] (31) at (3,1) {$0$};
    \node[state, diamond] (41) at (4,1) {$0$};
    \node[state, diamond] (02) at (0,2) {$0$};
    \node[state, diamond] (32) at (3,2) {$0$};
    \node[state, diamond] (42) at (4,2) {$0$};
    \node[state, rectangle, initial] (13) at (1,3) {$2$};
    \node[state, rectangle] (43) at (4,3) {$1$};
    \node[state, rectangle] (14) at (1,4) {$1$};
    \node[state, diamond] (24) at (2,4) {$0$};
    \node[state, diamond] (34) at (3,4) {$0$};
    \node[state, diamond] (44) at (4,4) {$0$};
    \node[state, diamond] (25) at (2,5) {$0$};
    \node[state, diamond] (35) at (3,5) {$0$};
    \node[below] at (00.south) {$( s_2,\langle \ins\rangle \mu Y. \phi_2)$};
    \node[below] at (10.south) {$\hspace*{3em}( s_2,\langle \std\rangle \phi)$};
    \node[below] at (40.south) {$\hspace*{-3em}( s_1,\langle \xxl\rangle \mu
      Y. \phi_2)$};
    \node[left] at (01.west) {$( s_2, \phi_3)$};
    \node[above] at (11.north) {$( s_2, \phi_2)$};
    \node[above] at (21.north) {$( s_2, \mu Y. \phi_2)$};
    \node[below] at (31.south) {$( s_1,\langle \ins\rangle \mu Y. \phi_2)$};
    \node[right] at (41.east) {$( s_1, \phi_3)$};
    \node[right] at (02.east) {$( s_2,\langle \xxl\rangle \mu Y. \phi_2)$};
    \node[below] at (32.south) {$( s_1,\langle \std\rangle \phi)$};
    \node[right] at (42.east) {$( s_1, \phi_2)$};
    \node[below] at (13.south) {$( s_0, \phi)$};
    \node[left] at (43.west) {$( s_1, \mu Y. \phi_2)$};
    \node[above] at (14.north) {$( s_0, \mu Y. \phi_2)$};
    \node[below] at (24.south) {$( s_0, \phi_2)$};
    \node[below] at (34.south) {$( s_0, \phi_3)$};
    \node[above] at (44.north) {$( s_0,\langle \ins\rangle \mu Y. \phi_2)$};
    \node[above] at (25.north) {$\hspace*{-1.5em}( s_0,\langle \std\rangle \phi)$};
    \node[above] at (35.north) {$\hspace*{1.5em}( s_0,\langle \xxl\rangle \mu Y. \phi_2)$};
    \path (00) edge[loop left] node[left] {$\ltrue$} (00);
    \path (10) edge[loop right] node[right] {$\ltrue$} (10);
    \path (40) edge[loop left] node[left] {$\ltrue$} (40);
    \path (25) edge[loop left] node[left] {$\ltrue$} (25);
    \path (35) edge[loop right] node[right] {$\ltrue$} (35);
    \path (31) edge[loop above] node[right] {$\neg \dollar$} (31);
    \path (32) edge[loop left] node[left] {$\neg \euro$} (32);
    \path (01) edge node[left] {$\ltrue$} (00);
    \path (11) edge node[left] {$\ltrue$} (10);
    \path (41) edge node[right] {$\ltrue$} (40);
    \path (42) edge node[right] {$\ltrue$} (41);
    \path (43) edge node[right] {$\ltrue$} (42);
    \path (44) edge node[right] {$\ltrue$} (43);
    \path (11) edge node[above] {$\ltrue$} (01);
    \path (21) edge node[below] {$\ltrue$} (11);
    \path (31) edge node[below] {$\dollar$} (21);
    \path (41) edge node[above] {$\ltrue$} (31);
    \path (01) edge node[left] {$\ltrue$} (02);
    \path (42) edge node[above] {$\ltrue$} (32);
    \path (32) edge[out=90, in=0] node[right] {$\;\;\euro$} (13);
    \path (13) edge node[left] {$\ltrue$} (14);
    \path (02) edge[out=90, in=180] node[left] {$\ltrue$} (14);
    \path (14) edge node[below] {$\ltrue$} (24);
    \path (24) edge node[below] {$\ltrue$} (34);
    \path (34) edge node[below] {$\ltrue$} (44);
    \path (24) edge node[right] {$\ltrue$} (25);
    \path (34) edge node[right] {$\ltrue$} (35);
  \end{tikzpicture}
  \Description{Featured parity game for checking whether $S\models \phi$.}
  \caption{Featured parity game for checking whether $S\models \phi$.}
  \label{fi:beek-pg}
\end{figure}
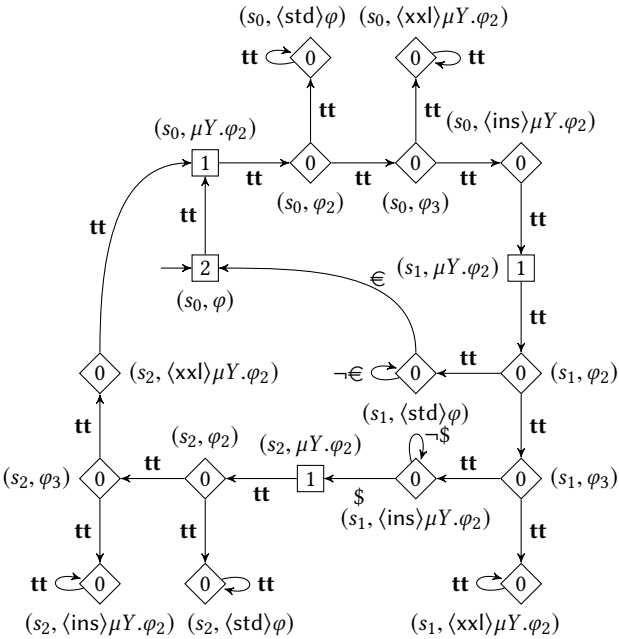

We show the result of the translation applied to our example in
Fig.~\ref{fi:beek-pg}, depicting only the reachable part of the
featured parity game.  Here, diamond-shaped states are owned by
player~1 and box-shaped states by player~2, and the priorities are
indicated inside states.  Player~1 is said to \emph{win} the game if
she can enforce an infinite path through the game graph for which the
highest priority occurring infinitely often is \emph{even}.  By the
properties of the translation~\cite{DBLP:conf/fase/BeekLVW20},
player~1 wins the game for a product $p$ iff the projection
$\proj{ p}{ S}$ satisfies $\phi$; in our case iff $\euro\in p$.
\cite{DBLP:conf/fase/BeekLVW20}~gives a family-based algorithm for
solving featured parity games.

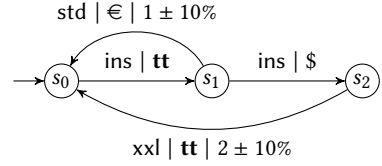
\begin{figure}[tbp]
  \centering
  \begin{tikzpicture}[x=2cm]
    \node[state, initial] (0) at (0,0) {$s_0$};
    \node[state] (1) at (1,0) {$s_1$};
    \node[state] (2) at (2,0) {$s_2$};
    \path (0) edge node[above] {$\ins\mid \ltrue$} (1);
    \path (1) edge node[above] {$\ins\mid \dollar$} (2);
    \path (1) edge[out=120, in=60] node[above] {$\std\mid \euro\mid
      1\pm 10\%$} (0);
    \path (2) edge[out=210, in=330] node[below] {$\xxl\mid \ltrue\mid
      2\pm 10\%$} (0);
  \end{tikzpicture}
  \Description{Coffee machine model $S$ with energy annotations.}
  \caption{Coffee machine model $S$ with energy annotations.}
  \label{fi:disc-fts}
\end{figure}

For another example of the use of games, we turn to the
\emph{quantitative} setting.  Figure~\ref{fi:disc-fts} displays our
toy model of the coffee machine together with approximate annotations
for energy consumption: brewing a standard coffee consumes $1$ energy
unit, plus/minus $10\%$; brewing an extra large coffee consumes
$2\pm 10\%$ energy units.  (Quite naturally, inserting coins does not
consume energy.)

We may now inquire about the \emph{robustness} of this SPL: given that
the energy annotations are approximate, what are the \emph{long-run}
deviations in energy consumption that we should expect, depending on
the particular product?  As a simple example, one machine might always
consume $1.1$ energy units at a $\std$ transition and another always
$0.9$, so that in an infinite run $( \ins, \std, \ins, \std,\dotsc)$
the two machines would accumulate a difference in energy consumption
of $0.2$ every second step.

Taking the standard point of view that \emph{the future is
  discounted}, we fix a discounting factor $\lambda< 1$ and multiply
differences by $\lambda$ at each step.  For the two runs above, the
long-run energy difference would thus evaluate to $0+ \lambda\cdot 0.2+
\lambda^2\cdot 0+ \lambda^3\cdot 0.2+\dotsm= 0.2\frac{ \lambda}{ 1-
  \lambda^2}$, which becomes $9.95$ for a standard discounting factor
of $\lambda= 0.99$.

Following~\cite{DBLP:journals/tcs/FahrenbergL14}, robustness of a
model for a product $p$ may be computed using the
\emph{$\lambda$-discounted bisimulation distance}: let $S_1$ and $S_2$
be the versions of the projection $\proj{ p}{ S}$ with the minimal,
resp.\ maximal, energy consumption on every transition and write
$S_i=\{ s_0^i, s_1^i, s_2^i\}$ for $i\in\{ 1, 2\}$, then the
discounted bisimulation distance between $S_1$ and $S_2$ is
$d( s_0^1, s_0^2)$, where $d: S_1\times S_2\to \Real$ is the unique
solution to the equation system given by
\begin{equation*}
  d( s^1, s^2)= \max
  \begin{cases}
    \max_{ s^1\ttto{ x}{ a} t^1} \min_{ s^2\ttto{ y}{ a} t^2}| x- y|+
    \lambda d( t^1, t^2) \\
    \max_{ s^2\ttto{ y}{ a} t^2} \min_{ s^1\ttto{ x}{ a} t^1}| x- y|+
    \lambda d( t^1, t^2)
  \end{cases}
\end{equation*}
for all $s^1\in S_1$, $s^2\in S_2$.  (Here $s\ttto{ x}{ a} t$
indicates a transition from $s$ to $t$ with label $a$ and energy
consumption $x$.)

\begin{figure}[t]
  \centering
  \begin{tikzpicture}[x=1.5cm, y=1.6cm]
    \node[state, diamond, initial] (0) at (0,0) {};
    \node[above left] at (0.north west) {$( s_0^1, s_0^2)$};
    \node[state, rectangle] (1) at (1,0) {};
    \node[below] at (1.south) {$( s_1^1, s_0^2, \ins, 0, 1)\;\;$};
    \node[state, diamond] (2) at (2,0) {};
    \node[below] at (2.south) {$( s_1^1, s_1^2)$};
    \node[state, rectangle] (3) at (3,0) {};
    \node[below] at (3.south) {$\;\;( s_2^1, s_1^2, \ins, 0, 1)$};
    \node[state, diamond] (4) at (4,0) {};
    \node[above] at (4.north) {$( s_2^1, s_2^2)$};
    \node[state, rectangle] (5) at (1,.8) {};
    \node[below] at (5.south) {$( s_0^1, s_1^2, \std, 0.9, 1)$};
    \node[state, rectangle] (6) at (2,-1.1) {};
    \node[above] at (6.north) {$( s_0^1, s_2^2, \xxl, 1.8, 1)$};
    \path (0) edge node[above] {$\ltrue\mid 0$} (1);
    \path (1) edge node[above] {$\ltrue\mid 0$} (2);
    \path (2) edge node[above] {$\dollar\mid 0$} (3);
    \path (3) edge node[above] {$\dollar\mid 0$} (4);
    \path (2) edge[out=90, in=0] node[right] {$\;\euro\mid 0$} (5);
    \path (5) edge[out=180, in=90] node[left] {$\euro\mid 0.2 \lambda^{ -1/2}$}
    (0);
    \path (4) edge[out=270, in=0] node[right] {$\;\;\ltrue\mid 0$} (6);
    \path (6) edge[out=180, in=270] node[left] {$\ltrue\mid 0.4 \lambda^{ -1/2}\;\;\;$}
    (0);
  \end{tikzpicture}
  \Description{Game for computing discounted distance.}
  \caption{Game for computing discounted distance.}
  \label{fi:disc-game}
\end{figure}
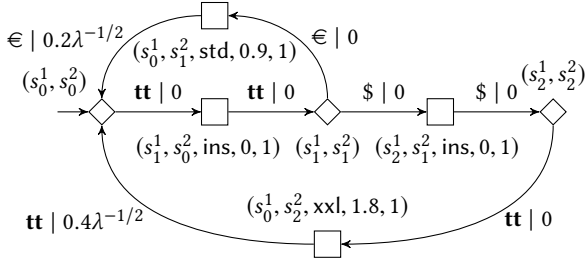

In~\cite{DBLP:conf/ictac/FahrenbergLQ19} it is shown that
$\lambda$-discounted bisimulation distances may be computed by
translating to $\sqrt{ \lambda}$-discounted
games~\cite{DBLP:journals/tcs/ZwickP96}.  We recall the translation
and extend it to FTS.  Let $F_1=( S_1, i_1, T_1, \gamma_1)$, $F_2=(
S_2, i_2, T_2, \gamma_2)$ be weighted FTS, with transitions
$T_j\subseteq S_j\times \Sigma\times \Rat\times S_j$.  The states of
the game for computing the $\lambda$-discounted bisimulation distance
between $F_1$ and $F_2$ are $V_1= S_1\times S_2$ (owned by player~1)
and $V_2= S_1\times S_2\times \Sigma\times \Rat\times\{ 1, 2\}$, with initial
state $i=( i_1, i_2)\in V_1$.  The transitions of the game are of four
types:
\begin{gather*}
  \{( s_1, s_2)\ttto{ \phi}{ 0}( s_1', s_2, a, x, 1)\mid( s_1, a, x,
  s_1')_{ / \phi}\in T_1\} \\
  \{( s_1, s_2)\ttto{ \phi}{ 0}( s_1, s_2', b, x, 2)\mid( s_2, b, x,
  s_2')_{ / \phi}\in T_2\} \\
  \{( s_1', s_2, a, x, 1)\ttto{ \phi}{ \lambda^{ -1/2}| a-
    b|}( s_1', s_2')\mid( s_2, b, x, s_2')_{ / \phi}\in T_2\} \\
  \{( s_1, s_2', b, x, 2)\ttto{ \phi}{ \lambda^{ -1/2}| a-
    b|}( s_1', s_2')\mid( s_1, a, x, s_1')_{ / \phi}\in T_1\}
\end{gather*}
We show the result of the translation applied to our example in
Fig.~\ref{fi:disc-game}, where we have omitted some states and
transitions due to symmetry.  For $\lambda= 0.99$ and
$p=\{ \euro, \dollar\}$, the distance evaluates to $13.2$.

Games are also important in \emph{controller synthesis}: the problem
of generating controllers for discrete event
systems~\cite{journals/pieee/RamadgeW89, book/KumarG12}.  In this
setting, the model is a game in which player~1 is the
\emph{controller} and player~2 the \emph{environment}, and then the
task is to find a \emph{strategy} for the controller which ensures a
given property one wishes to enforce.

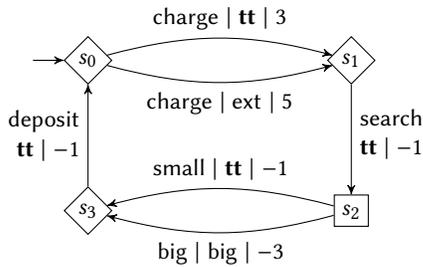
\begin{figure}[b]
  \centering
  \begin{tikzpicture}[x=3.5cm, y=2cm]
    \node[state, diamond, initial] (0) at (0,0) {$s_0$};
    \node[state, diamond] (1) at (1,0) {$s_1$};
    \node[state, rectangle] (2) at (1,-1) {$\,s_2\,$\raisebox{-.3em}{\rule{0pt}{1em}}};
    \node[state, diamond] (3) at (0,-1) {$s_3$};
    \path (0) edge[out=15, in=165] node[above] {$\charge\mid
      \ltrue\mid 3$} (1);
    \path (0) edge[out=-15, in=-165] node[below] {$\charge\mid
      \fextra\mid 5$} (1);
    \path (1) edge node[right, align=left, pos=.45] {$\search$ \\ $\ltrue\mid
      -1$} (2);
    \path (2) edge[out=165, in=15] node[above] {$\srock\mid
      \ltrue\mid -1$} (3);
    \path (2) edge[out=-165, in=-15] node[below] {$\brock\mid
      \fbrock\mid -3$} (3);
    \path (3) edge node[left, align=right] {$\deposit$ \\ $\ltrue\mid
      -1$} (0);
  \end{tikzpicture}
  \Description{A simple energy game.}
  \caption{A simple energy game.}
  \label{fi:egame}
\end{figure}

We give a simple example in Fig.~\ref{fi:egame}, inspired
by~\cite{DBLP:conf/tase/BauerJLSL12}.  This is a toy model of a mars
robot which collects rocks, with an operations cycle consisting of
charging its batteries, searching for rocks, collecting a rock, and
deposing the rock in a container.  Charging the battery adds $3$
energy units to its battery; unless the $\fextra$ feature is present,
in which case the $\charge$ transition may add $5$ energy units.
Searching and deposing both cost $1$ energy unit, as does collecting a
small rock.  If the $\fbrock$ feature is present, then also big rocks
may be collected, with an energy consumption of $3$.  The size of a
collected rock is controlled by the environment.

The property we wish to enforce is that the system have an infinite
run in which the battery charge never drops below $0$.  That is,
player~1 should have a strategy of choosing her transitions so that no
matter the behavior of player~2, battery charge never goes negative.
A simple analysis shows that this is the case precisely for all
products which satisfy the formula $\neg \fbrock\lor \fextra$: if
feature $\fbrock$ is not present, then the search-collect-deposit
cycle always consumes $3$ energy units which can be recharged also
without the $\fextra$ feature; and if both $\fbrock$ and $\fextra$ are
present, then charging $5$ energy units ensures that also big rocks
can be deposited.

In this paper we concern ourselves with several types of games which
have been used in model checking and controller synthesis.  We lift
these games to featured versions useful in an SPL context, and we show
how to compute their values and optimal strategies in a family-based
manner using late splitting~\cite{DBLP:conf/icse/ApelRWGB13}.  We
treat the following types of games:
\begin{itemize}
\item reachability games;
\item minimum reachability games;
\item discounted games;
\item energy games;
\item parity games.
\end{itemize}

Our treatment is based on the computation of \emph{attractors}, which
in general is the most efficient technique for solving games and
typically gives raise to (pseudo)polynomial algorithms.  Our first
main contribution is showing how to lift attractor computations to the
featured setting, in Sections~\ref{se:reach} through~\ref{se:parity}.
(Compared to~\cite{DBLP:conf/fase/BeekLVW20}, we use a different
algorithm for parity games which is known to be more
efficient~\cite{DBLP:conf/tacas/Dijk18}.)

Our second main contribution, in Section~\ref{se:opt}, is the
family-based computation of optimal strategies.  We show that in all
featured games considered here, optimal featured strategies may be
found during the attractor computation, and these project to optimal
strategies for \emph{any} product.

Finally, Section~\ref{se:comp} exhibits our third main contribution:
family-based algorithms, using late splitting, to compute attractors
for all the featured games we have introduced. %
\iflongversion%
\nocite{journals/corr/FahrenbergL20}
\fi

\section{Featured Reachability Games}
\label{se:reach}

\subsection{Reachability Games}

A \emph{game structure} $G=( S_1, S_2, i, F, T)$ consists of two
disjoint sets $S= S_1\sqcup S_2$ of \emph{states}, \emph{initial} and
\emph{accepting states} $i\in S$, $F\subseteq S$, and
\emph{transitions} $T\subseteq S\times S$.  For simplicity we assume
$G$ to be \emph{non-blocking}, so that for all $s\in S$ there exists
$s'\in S$ for which $( s, s')\in T$.

As customary, we write $s\to s'$ to indicate that $( s, s')\in T$.
Intuitively, a game on a game structure $G$ as above is played by two
players, player~1 and player~2, taking turns to move a token along the
directed graph with vertices $S$ and edges $T$.  We to make this
intuition precise.

A \emph{finite path} in $G$ is a finite sequence
$\pi=( s_1,\dotsc, s_k)$ in $S$ such that $s_i\to s_{ i+ 1}$ for all
$i= 1,\dotsc, k- 1$.  The set of finite paths in $G$ is denoted
$\fpaths{ G}$.  The \emph{end state} of a path
$\pi=( s_1,\dotsc, s_k)$ is $\pend{ \pi}= s_k$.  An \emph{infinite
  path} in $G$ is an infinite sequence $( s_1, s_2,\dotsc)$ in $S$
such that $s_i\to s_{ i+ 1}$ for all $i\ge 1$, and the set of these is
denoted $\ipaths{ G}$.

The \emph{configurations} for player~$i$, for $i\in\{ 1, 2\}$, are
$\Conf_i=\{ \pi\in \fpaths{ G}\mid \pend{ \pi}\in S_i\}$.  A
\emph{strategy} for player~$i$ is a function $\theta: \Conf_i\to S$
such that for all $\pi\in \Conf_i$, $\pend{ \pi}\to \theta( \pi)$.
The set of strategies for player~$i$ is denoted $\Theta_i$.

Any pair of strategies $\theta_1\in \Theta_1$, $\theta_2\in \Theta_2$
induces a unique infinite path
$\out{ \theta_1}{ \theta_2}=( s_1, s_2,\dotsc)\in \ipaths{ G}$, called
the \emph{outcome} of the pair $( \theta_1, \theta_2)$ and defined
inductively as follows:
\begin{equation*}
  s_1= i \qquad s_{ 2 k}= \theta_1( s_1,\dotsc, s_{ 2 k- 1}) \qquad
  s_{ 2 k+ 1}= \theta_2( s_1,\dotsc, s_{ 2 k})
\end{equation*}
Note that the outcome is indeed infinite due to our non-blocking
assumption.

The \emph{reachability game} on a game structure
$G=( S_1, S_2, i, F, T)$ is to decide whether there exists a strategy
$\theta_1\in \Theta_1$ such that for all $\theta_2\in \Theta_2$,
writing $\out{ \theta_1}{ \theta_2}=( s_1, s_2,\dotsc)$, there is an
index $k\ge 1$ for which $s_k\in F$.  In the affirmative case,
player~1 is said to \emph{win} the reachability game.

In order to solve reachability games, one introduces a notion of
\emph{game attractor} $\attr:( S\to \Bool)\to( S\to \Bool)$, where
$\Bool=\{ \lfalse, \ltrue\}$ is the boolean lattice, defined for any
$U: S\to \Bool$ by
\begin{equation*}
  \attr( U)( s)=
  \begin{cases}
    \bigvee_{ s\to s'} U( s') &\text{if } s\in S_1\,, \\
    \bigwedge_{ s\to s'} U( s') &\text{if } s\in S_2\,.
  \end{cases}
\end{equation*}
Hence $\attr( U)( s)$ is true precisely if \emph{there exists} a
player-1 transition to a state $s'$ for which $U( s')= \ltrue$, or if
it holds \emph{for all} player-2 transitions $s\to s'$ that
$U( s')= \ltrue$.

Let $\attr^*= \id\lor \attr\lor \attr^2\lor\dotsm$, where $\lor$ is
the supremum operator on the complete lattice $S\to \Bool$.  The
following is then easy to see.

\begin{lemma}
  Let $G=( S_1, S_2, i, F, T)$ be a game structure and define
  $I: S\to \Bool$ by $I( s)= \ltrue$ iff $s\in F$.  Player~1 wins the
  reachability game in $G$ iff $\attr^*( I)( i)= \ltrue$.
\end{lemma}

%

The operator $\attr$ is monotone on the complete lattice $S\to \Bool$,
thus $\attr^*$ can be computed using a fixed-point algorithm, in time
quadratic in the size of $S$.  Hence reachability games can be decided
in polynomial time.

\subsection{Featured Reachability Games}

Let $N$ be a finite set of \emph{features} and $\px\subseteq 2^N$ a
set of \emph{products} over $N$.  A \emph{feature guard} is a Boolean
expression over $N$, and we denote the set of these by $\Bool( N)$.
We write $p\models \gamma$ if $p\in \px$ satisfies
$\gamma\in \Bool( N)$.
For each $p\in \px$ let $\gamma_p\in \Bool( N)$ be its
\emph{characteristic formula} satisfying that $p'\models \gamma_p$ iff
$p'= p$.

A \emph{featured game structure} $G=( S_1, S_2, i, F, T, \gamma)$
consists of a game structure $( S_1, S_2, i, F, T)$ together with a
mapping $\gamma: T\to \Bool( N)$.  We also assume our featured game
structures to be non-blocking, in the sense that for all $s\in S$ and
all $p\in \px$, there exists $( s, s')\in T$ with $p\models \gamma( s,
s')$.

The \emph{projection} of a featured game structure $G$ as above to a
product $p\in \px$ is the game structure $\proj{ p}{ G}=( S_1, S_2, i,
F, T')$ with $T'=\{ t\in T\mid p\models \gamma( t)\}$.  All
projections of non-blocking featured game structures are again
non-blocking.

We are interested in solving the reachability game for each product
$p\in \px$, but in a family-based manner.  We will thus compute a
function $\Bool( N)\to \Bool$ which for each feature expression
indicates whether player~1 wins the reachability game on $G$.

To this end, define the \emph{featured attractor}
$\fattr:( S\to( \Bool( N)\to \Bool))\to( S\to( \Bool( N)\to \Bool))$
by
\begin{equation*}
  \fattr( U)( s)( \phi)=
  \begin{cases}
    \bigvee_{ s\to s'} U( s')( \gamma(( s, s'))\land \phi) &\text{if }
    s\in S_1\,, \\
    \bigwedge_{ s\to s'} U( s')( \gamma(( s, s'))\land \phi) &\text{if
    } s\in S_2\,.
  \end{cases}
\end{equation*}
and let $\fattr^*= \id\vee \fattr\vee \fattr^2\vee \dotsm$, the
supremum in the complete lattice $S\to( \Bool( N)\to \Bool)$.


\begin{theorem}
  \label{th:fattr}
  Let $G=( S_1, S_2, i, F, T, \gamma)$ be a featured game structure
  and define $I: S\to( \Bool( N)\to \Bool)$ by $I( s)( \phi)= \ltrue$
  if $s\in F$; $\lfalse$ if $s\notin F$.  Let $p\in \px$, then
  Player~1 wins the reachability game in $\proj{ p}{ G}$ iff
  $\fattr^*( I)( i)( \gamma_p)= \ltrue$.
\end{theorem}

The operator $\fattr$ is monotone on the complete lattice
$S\to( \Bool( N)\to \Bool)$, thus $\fattr^*$ can be computed using a
fixed-point algorithm.  In Section~\ref{se:comp} we will give an
algorithm which uses \emph{guard
  partitions}~\cite{DBLP:journals/sttt/FahrenbergL19} and \emph{late
  splitting}~\cite{DBLP:conf/icse/ApelRWGB13} to compute the fixed
point.

\section{Featured Minimum Reachability}
\label{se:minreach}

We now enrich the above problem to compute featured \emph{minimum}
reachability in \emph{weighted} game structures.

\subsection{Minimum Reachability Games}

A \emph{weighted game structure} $G=( S_1, S_2, i, F, T)$ consists of
two disjoint sets $S= S_1\sqcup S_2$ of states, initial and accepting
states $i\in S$, $F\subseteq S$, and transitions
$T\subseteq S\times \Nat\times S$.  Note that all weights are
non-negative.  We also assume our weighted game structures to be
non-blocking, and we write $s\tto{ x} s'$ to indicate that
$( s, x, s')\in T$.

Games on such structures are played as before, only now the goal of
player~1 is not only to reach a state in $F$, but to do so as cheaply
as possible.  Let us make this precise.  A path in $G$ is now a
sequence $\pi=( s_1, x_1, s_2, x_2,\dotsc)$ such that
$s_i\tto{ x_i} s_{ i+ 1}$ for all $i= 1$.  The notion of configuration
is unchanged, and a strategy for player~$i$ is now a function
$\theta: \Conf_i\to \Nat\times S$ such that for all
$\pi\in \Conf_i$, $\pend{ \pi}\tto{ \theta( \pi)_1} \theta( \pi)_2$,
where $\theta( \pi)=( \theta( \pi)_1, \theta( \pi)_2)$.  The outcome
of a strategy pair is an infinite path
$( s_1, x_1, s_2, x_2,\dotsc)\in \ipaths{ G}$ defined as expected.

The \emph{reachability value} of an infinite path
$\pi=( s_1, x_1, s_2, x_2,\dotsc)$ is defined to be
$\valr( \pi)=\min\{ x_1+\dotsm+ x_{ k- 1}\mid s_k\in F\}$, where
$\min \emptyset= \infty$ by convention, and the \emph{minimum
  reachability value}
\linebreak 
of $G$ is
$\valr( G)= \inf_{ \theta_1\in \Theta_1} \sup_{ \theta_2\in \Theta_2}
\valr( \out{ \theta_1}{ \theta_2})$.  That is, %
\linebreak 
$\valr( \out{ \theta_1}{ \theta_2})$ is the minimum sum of weights
along any accepting finite path, and the goal of player~1 is to
minimize this value.

In order to compute minimum reachability in a weighted game structure
$G$, define the \emph{weighted attractor}
$\wattr:( S\to \Nat\cup\{ \infty\})\to( S\to \Nat\cup\{ \infty\})$ by
\begin{align*}
  \wattr( U)( s)=
  \begin{cases}
    \min_{ s\tto{ x} s'} x+ U( s') &\text{if } s\in S_1\,, \\
    \max_{ s\tto{ x} s'} x+ U( s') &\text{if } s\in S_2
  \end{cases}
\end{align*}
and let $\wattr^*= \min(\id, \wattr,$ $\wattr^2,\dotsc)$.
The following seems to be folklore; note that it only holds under our
assumption that all weights are non-negative.
(See~\cite{DBLP:journals/acta/BrihayeGHM17} for an extension to
negative weights.)

\begin{lemma}
  The minimum reachability value of a weighted game structure
  $G=( S_1, S_2, i, F, T)$ is $\valr( G)= \wattr^*( I)( i)$, where
  $I: S\to \Nat\cup\{ \infty\}$ is defined by $I( s)= 0$ if
  $s\in F$; $\infty$ if $s\notin F$.
\end{lemma}

The operator $\wattr$ is monotone on the complete lattice of functions
$S\to \Nat\cup\{ \infty\}$, thus $\wattr^*$ can be computed using a
fixed-point algorithm, in time quadratic in the size of $S$ and
linear in the maximum of the weights on the transitions of $G$.  That
is, minimum reachability values can be computed in pseudo-polynomial
time.

\subsection{Featured Minimum Reachability Games}

A \emph{featured weighted game structure}
$G=( S_1, S_2, i, F, T, \gamma)$ consists of a weighted game structure
$( S_1, S_2, i, F, T)$ together with a mapping
$\gamma: T\to \Bool( N)$.  We again assume our featured weighted game
structures to be non-blocking.  Projections of such structures to
products $p\in \px$ are defined as before.

Define the \emph{featured weighted attractor} operator
$\fwattr:( S\to( \Bool( N)\to \Nat\cup\{ \infty\}))\to( S\to( \Bool(
N)\to \Nat\cup\{ \infty\}))$ by
\begin{align*}
  \fwattr( U)( s)( \phi)=
  \begin{cases}
    \min_{ s\tto{ x} s'} x+ U( s')( \gamma(( s, x, s'))\land \phi)
    &\text{if } s\in S_1, \\
    \max_{ s\tto{ x} s'} x+ U( s')( \gamma(( s, x, s'))\land \phi)
    &\text{if } s\in S_2
  \end{cases}
\end{align*}
and let $\fwattr^*= \min( \id, \fwattr, \fwattr^2,\dotsc)$.

\begin{theorem}
  \label{th:fwattr}
  Let $G=( S_1, S_2, i, F, T, \gamma)$ be a featured weighted game
  structure and define $I: S\to( \Bool( N)\to \Nat\cup\{ \infty\})$ by
  $I( s)( \phi)= 0$ if $s\in F$; $\infty$ if $s\notin F$.  Let
  $p\in \px$, then the minimum reachability value of $\proj{ p}{ G}$
  is $\valr( \proj{ p}{ G})= \fwattr^*( I)( i)( \gamma_p)$.
\end{theorem}

\section{Featured Discounted Games}
\label{se:disc}

\subsection{Discounted Games}

We now turn our attention towards \emph{discounted} games.  These are
also played on weighted game structures, but now the accepting states
are ignored, and the restriction on non-negativity of weights can be
lifted.  That is, we are now working with weighted game structures
$G=( S_1, S_2, i, T)$ with $T\subseteq S\times \Int\times S$.

The notions of configurations, strategies, and outcome remain
unchanged from the previous section.  Let $0< \lambda< 1$ be a real
number, called the \emph{discounting factor} of the game.  The
\emph{discounted value} of an infinite path
$\pi=( s_1, x_1, s_2, x_2,\dotsc)$ is
$\vall{ \lambda}( \pi)= x_1+ \lambda x_2+ \lambda^2 x_3+\dotsm$, and
the discounted value of a game $G$ is
$\vall{ \lambda}( G)= \sup_{ \theta_1\in \Theta_1} \inf_{ \theta_2\in
  \Theta_2} \vall{ \lambda}( \out{ \theta_1}{ \theta_2})$.  That is,
the value of a path is the sum of its weights, progressively
discounted along its run, and the goal of player~1 is to maximize this
value.

The following is a reformulation of a result
from~\cite{DBLP:journals/tcs/ZwickP96} in terms of attractors.  Define
the \emph{discounted attractor} $\dattr:( S\to \Real)\to( S\to \Real)$
by
\begin{align*}
  \dattr( U)( s)=
  \begin{cases}
    \max_{ s\tto{ x} s'} x+ \lambda U( s') &\text{if } s\in S_1\,, \\
    \min_{ s\tto{ x} s'} x+ \lambda U( s') &\text{if } s\in S_2\,.
  \end{cases}
\end{align*}

\begin{lemma}
  Let $G=( S_1, S_2, i, T)$ be a weighted game structure.  The
  equation system $V= \dattr( V)$ has a unique solution $\dattr^*$,
  and the discounted value of $G$ is
  $\vall{ \lambda}(G)= \dattr^*( i)$.
\end{lemma}

\subsection{Featured Discounted Games}

Let $G=( S_1, S_2, i, T, \gamma)$ be a featured weighted game
structure.  Define the \emph{featured discounted attractor}
$\fdattr:( S\to( \Bool( N)\to \Real))\to( S\to( \Bool( N)\to \Real))$
by
\begin{equation*}
  \fdattr( U)( s)( \phi)=
  \begin{cases}
    \max_{ s\tto{ x} s'} x+ \lambda U( s')( \gamma(( s, x, s'))\land
    \phi) &\text{if } s\in S_1, \\
    \min_{ s\tto{ x} s'} x+ \lambda U( s')( \gamma(( s, x, s'))\land
    \phi) &\text{if } s\in S_2.
  \end{cases}
\end{equation*}

\begin{theorem}
  \label{th:fdattr}
  Let $G=( S_1, S_2, i, T, \gamma)$ be a featured weighted game
  structure.  The equation system $V= \fdattr( V)$ has a unique
  solution $\fdattr^*$, and for any $p\in \px$, the discounted value
  of $\proj{ p}{ G}$ is
  $\vall{ \lambda}( \proj{ p}{ G})= \fdattr^*( i)( \gamma_p)$.
\end{theorem}

\paragraph{Example}

We show the computation of $\fdattr^*$ for the example from
Fig.~\ref{fi:disc-game}; recall that this is a
$\sqrt{ \lambda}$-discounted game.  For any $\phi\in \Bool( N)$, and
writing $i=(( s^1_0, s^2_0))$, we
have
\begin{align*}
  \fdattr^*( i)( \phi) &= \sqrt{ \lambda} \fdattr^*((
  s^1_1, s^2_0, \ins))( \phi) \\
  &= \sqrt{ \lambda}^2 \fdattr^*(( s^1_1, s^2_1))( \phi)
  \intertext{and, skipping computations for player-2 states from now,}
  &= \max
  \begin{cases}
    \sqrt{ \lambda}^3\cdot 0.2\tfrac{ 1}{ \sqrt{ \lambda}}+ \sqrt{
      \lambda}^4 \fdattr^*( i)( \phi\land \euro) \\
    \sqrt{ \lambda}^4 \fdattr^*(( s^1_2, s^2_2))( \phi\land \dollar)
  \end{cases} \\
  &= \max
  \begin{cases}
    \sqrt{ \lambda}^3\cdot 0.2\tfrac{ 1}{ \sqrt{ \lambda}}+ \sqrt{
      \lambda}^4 \fdattr^*( i)( \phi\land \euro) \\
    \sqrt{ \lambda}^5\cdot 0.4\tfrac{ 1}{ \sqrt{ \lambda}}+ \sqrt{
      \lambda}^6 \fdattr^*( i)( \phi\land \dollar)
  \end{cases} \\
  &= \max
  \begin{cases}
    0.2\lambda+ \lambda^2 \fdattr^*( i)( \phi\land \euro)
    \\
    0.4\lambda^2+ \lambda^3 \fdattr^*( i)( \phi\land
    \dollar)
  \end{cases}
\end{align*}

For $p=\{ \euro\}$, we have
$\fdattr^*( i)( \gamma_{\{ \euro\}})= 0.2\lambda+ \lambda^2 \fdattr^*(
i)( \gamma_{\{ \euro\}})$, hence
$\vall{ \lambda}( \proj{\{ \euro\}}{ G})= 0.2 \frac{ \lambda}{ 1-
  \lambda^2}$.  Given that $( \ins, \std)^\omega$ is the only infinite
run in the projection $\proj{\{ \euro\}}{ S}$ of the original model,
this is as expected.

For $p=\{ \dollar\}$, the equation simplifies to
$\fdattr^*( i)( \gamma_{\{ \dollar\}})= 0.4\lambda^2+ \lambda^3
\fdattr^*( i)( \gamma_{\{ \dollar\}})$, hence
$\vall{ \lambda}( \proj{ \{ \dollar\}}{ G})= 0.4 \frac{ \lambda^2}{ 1-
  \lambda^3}$.  For $p=\{ \euro, \dollar\}$, no simplifications are
possible; for $\lambda= 0.99$ a standard fixed-point iteration yields
$\vall{ \lambda}( \proj{ \{ \euro, \dollar\}}{ G})= 13.2$.

\section{Featured Energy Games}
\label{se:energy}

\subsection{Energy Games}

Energy games are played on the same type of weighted game structures
as the discounted games of the previous section, and also the notions
of configurations, strategies, and outcome remain unchanged.

Let $v_0\in \Nat$.  An infinite path
$( s_1, x_1, s_2, x_2,\dotsc)\in \ipaths{ G}$ in a weighted game
structure $G$ is \emph{energy positive with initial credit $v_0$} if
all finite sums $v_0+ x_1, v_0+ x_1+ x_2,\dotsc$ are non-negative;
that is, if $v_0+ \sum_{ i= 1}^k x_i\ge 0$ for all $k\ge 1$.  The
\emph{energy game} on $G$ with initial credit $v_0$ is to decide
whether there exists a strategy $\theta_1\in \Theta_1$ such that for
all $\theta_2\in \Theta_2$, $\out( \theta_1, \theta_2)$ is energy
positive with initial credit $v_0$.

The following procedure, first discovered
in~\cite{DBLP:journals/fmsd/BrimCDGR11}, can be used to solve energy
games.  Let $G=( S_1, S_2, i, T)$ be a weighted game structure and
define
$M= \sum_{ s\in S} \max(\{ 0\}\cup\{ -x\mid( s, x, s')\in T\})$.  Let
$W=\{ 0,\dotsc, M, \top\}$, where $\top$ is the greatest element, and
define an operation $\mathord{\ominus}: W\times \Int\to W$ by
$x\ominus y= \max( 0, x- y)$ if $x\ne \top$ and $x- y\le M$; $\top$
otherwise.

Now define the \emph{energy attractor} $\eattr:( S\to W)\to( S\to W)$
by
\begin{equation*}
  \eattr( U)( s)=
  \begin{cases}
    \min_{ s\tto{ x} s'} U( s')\ominus x &\text{if } s\in S_1\,, \\
    \max_{ s\tto{ x} s'} U( s')\ominus x &\text{if } s\in S_2
  \end{cases}
\end{equation*}
and let $\eattr^*= \max( \id, \eattr, \eattr^2,\dotsc)$.  The
following is proven in~\cite{DBLP:journals/fmsd/BrimCDGR11} which also
shows that energy games can be decided in pseudo-polynomial time.

\begin{lemma}
  Let $G=( S_1, S_2, i, T)$ be a weighted game structure and
  $v_0\in \Nat$.  Player~1 wins the energy game on $G$ with initial
  credit $v_0$ iff $v_0\ge \eattr^*( I)( i)$, where $I: S\to W$ is
  defined by $I( s)= 0$ for all $s\in S$.
\end{lemma}

\subsection{Featured Energy Games}

Let $G=( S_1, S_2, i, T, \gamma)$ be a featured weighted game
structure.  Define the \emph{featured energy attractor}
$\feattr:( S\to( \Bool( N)\to W))\to( S\to( \Bool( N)\to W))$ by
\begin{equation*}
  \feattr( U)( s)( \phi)=
  \begin{cases}
    \min_{ s\tto{ x} s'} U( s')( \gamma(( s, x, s'))\land \phi)\ominus
    x &\text{if } s\in S_1\,, \\
    \max_{ s\tto{ x} s'} U( s')( \gamma(( s, x, s'))\land \phi)\ominus
    x &\text{if } s\in S_2
  \end{cases}
\end{equation*}
and let $\feattr^*= \max( \id, \feattr, \feattr^2,\dotsc)$.

\begin{theorem}
  \label{th:feattr}
  Let $G=( S_1, S_2, i, T, \gamma)$ be a featured weighted game
  structure, $v_0: \Bool( N)\to \Nat$, and define
  $I: S\to( \Bool( N)\to W)$ by $I( s)( \phi)= 0$ for all $s\in S$,
  $\phi\in \Bool( N)$.  Let $p\in \px$, then player~1 wins the energy
  game on $\proj{ p}{ G}$ with initial credit $v_0( \gamma_p)$ iff
  $v_0( \gamma_p)\ge \feattr^*( I)( i)( \gamma_p)$.
\end{theorem}

\paragraph{Example}

We show the computation of $\fdattr^*$ for the example from
Fig.~\ref{fi:egame}.  Note that the example includes labels on
transitions; for energy computations, these are ignored.  We have $M=
3$ and thus $W=\{ 0, 1, 2, 3, \top\}$.  Denote $\feattr^*( I)= f^*$,
then for any $\phi\in \Bool( N)$,
\begin{align*}
  f^*( i)( \phi) &= \min
  \begin{cases}
    f^*( s_1)( \phi)\ominus 3 \\
    f^*( s_1)( \phi\land \fextra)\ominus 5
  \end{cases} \\
  &= \min
  \begin{cases}
    ( f^*( s_2)( \phi)\ominus -1)\ominus 3 \\
    ( f^*( s_2)( \phi\land \fextra)\ominus -1)\ominus 5
  \end{cases} \\
  &\hspace*{-2em}= \min
  \begin{cases}
    \max
    \begin{cases}
      (( f^*( s_3)( \phi)\ominus -1)\ominus -1)\ominus 3 \\
      (( f^*( s_3)( \phi\land \fbrock)\ominus -3)\ominus -1)\ominus 3
    \end{cases} \\
    \max
    \begin{cases}
      (( f^*( s_3)( \phi\land \fextra)\ominus -1)\ominus -1)\ominus 5 \\
      (( f^*( s_3)( \phi\land \fextra\land \fbrock)\ominus -3)\ominus -1)\ominus 5
    \end{cases}
  \end{cases} \\
  &\hspace*{-2em}= \min
  \begin{cases}
    \max
    \begin{cases}
      ((( f^*( i)( \phi)\ominus -1)\ominus -1)\ominus -1)\ominus 3 \\
      ((( f^*( i)( \phi\land \fbrock)\ominus -1)\ominus -3)\ominus -1)\ominus 3
    \end{cases} \\
    \max
    \begin{cases}
      ((( f^*( i)( \phi\land \fextra)\ominus -1)\ominus -1)\ominus -1)\ominus 5 \\
      ((( f^*( i)( \phi\land \fextra\land \fbrock)\ominus -1)\ominus -3)\ominus -1)\ominus 5
    \end{cases}
  \end{cases}
\end{align*}

For $p= \emptyset$, only the first of these four lines contributes to
the fixed point, which thus becomes $f^*( i)( \gamma_{ \emptyset})=
\feattr^*( I)( i)( \gamma_{ \emptyset})= 0$.  Hence the minimum
necessary initial credit in the energy game without any extra features
is $0$, as expected.  For the other three products, standard
fixed-point iterations yield $f^*( i)( \gamma_{\{ \fbrock\}})= \top$
(player~1 cannot win this game) and $f^*( i)( \gamma_{\{ \fextra\}})=
f^*( i)( \gamma_{\{ \fextra, \fbrock\}})= 0$.

\section{Featured Parity Games}
\label{se:parity}

\subsection{Parity Games}

A \emph{priority game structure} $G=( S_1, S_2, i, T, p)$ is a game
structure (without weights) together with a \emph{priority} mapping
$p: S\to \Nat$; we again assume these to be non-blocking.  The notions
of configurations, strategies and outcomes remain unchanged.

For an infinite path $\pi=( s_1, s_2,\dotsc)\in \ipaths{ G}$ let
$\prio{ \pi}= \liminf_{ n\to \infty} p(n)$ be the lowest priority
which occurs infinitely often in $\pi$.  The \emph{parity game} on $G$
is to decide whether there exists a strategy $\theta_1\in \Theta_1$
such that for all $\theta_2\in \Theta_2$,
$\prio{ \out{ \theta_1}{ \theta_2}}$ is an even number.

Note that this is, thus, a \emph{minimum} parity game, whereas the
game we exposed in the introduction was a \emph{maximum} parity game.
This unfortunate dissonance between model checking and game theory,
which we choose to embrace rather than fix here, can easily be
overcome by inverting all priorities and then adding their former
maximum.

The following procedure for solving minimum parity games was
discovered in~\cite{DBLP:conf/stacs/Jurdzinski00}.\uli{Check very
  carefully} Let $G=( S_1, S_2, i, T, p)$ be a priority game structure
and $d= \max\{ p( s)\mid s\in S\}$.  For every $i\in\{ 0,\dotsc, d\}$
let $p_i=|\{ s\in S\mid p( s)= i\}|$ be the number of states with
priority $i$ and define $M'\subseteq \Nat^d$ to be the following
(finite) set: if $d$ is odd, then
$M'=\{ 0\}\times\{ 0,\dotsc, p_1\}\times\{ 0\}\times\{ 0,\dotsc,
p_3\}\times\dotsm\times\{ 0,\dotsc, p_d\}$; if $d$ is even, then
$M'=\{ 0\}\times\{ 0,\dotsc, p_1\}\times\{ 0\}\times\{ 0,\dotsc,
p_3\}\times\dotsm\times\{ 0\}$.

We need some notation for lexicographic orders on $\Nat^d$.  For $x=(
x_1,\dotsc, x_d), y=( y_1,\dotsc, y_d)\in \Nat^d$ and $k\in\{
1,\dotsc, d\}$, say that $x\le_k y$ if $x_i\le y_i$ for all components
$i\in\{ 1,\dotsc, k\}$.  Relations $=_k$, $<_k$, $\ge_k$ and $>_k$ are
defined similarly.

Let $M= M'\cup\{ \top\}$, where $\top$ is the greatest element in all
the orders $\le_k$, and define the relations $\preceq_k$ on $M$ by
$x\preceq_k y$ iff $x\le_k y$ if $k$ is odd; $x<_k y$ or $x= y= \top$
if $k$ is even.  Define a function
$\prog:( S\to M)\times S\times S\to M$ by
$\prog( U, s, s')= \min\{ m\in M\mid m\succeq_{ p( s)+ 1} U( s')\}$.

Now define the \emph{parity attractor} $\pattr:( S\to M)\to( S\to M)$
by
\begin{equation*}
  \pattr( U)( s)=
  \begin{cases}
    \min_{ s\to s'} \prog( U, s, s') &\text{if } s\in S_1\,, \\
    \max_{ s\to s'} \prog( U, s, s') &\text{if } s\in S_2
  \end{cases}
\end{equation*}
and let $\pattr^*= \max( \id, \pattr, \pattr^2,\dotsc)$.  The
following is shown in~\cite{DBLP:conf/stacs/Jurdzinski00}, together
with the fact that parity games are decidable in pseudo-polynomial
time.

\begin{lemma}
  Let $G=( S_1, S_2, i, T, p)$ be a priority game structure and define
  $I: S\to M$ by $I( s)=( 0,\dotsc, 0)$ for all $s\in S$.  Player~1
  wins the parity game on $G$ iff $\pattr^*( I)( i)\ne \top$.
\end{lemma}

\subsection{Featured Parity Games}

A \emph{featured priority game structure}
$G=( S_1, S_2, i, T, p, \gamma)$ consists of a priority game structure
$G=( S_1, S_2, i, T, p)$ together with a mapping
$\gamma: T\to \Bool( N)$.  We again assume these to be non-blocking.
Let $d= \max\{ p( s)\mid s\in S\}$ and $M$ be defined as above.

Let
$\fprog:( S\to( \Bool( N)\to M))\times S\times S\to( \Bool( N)\to M)$
be the function given by
$\fprog( U, s, s')( \phi)= \min\{ m\in M\mid m\succeq_{ p( s)} U( s')(
\phi)\}$.  Define the \emph{featured parity attractor}
$\fpattr:( S\to( \Bool( N)$ $\to M))\to( S\to( \Bool( N)\to M))$ by
\begin{equation*}
  \fpattr( U)( s)( \phi)=
  \begin{cases}
    \min\limits_{ s\to s'} \fprog( U, s, s')( \gamma(( s, s'))\land
    \phi) &\text{if } s\in S_1\,, \\
    \max\limits_{ s\to s'} \fprog( U, s, s')( \gamma((s, s'))\land
    \phi) &\text{if } s\in S_2
  \end{cases}
\end{equation*}
and let $\fpattr^*= \max( \id, \fpattr, \fpattr^2,\dotsc)$.

\begin{theorem}
  \label{th:fpattr}
  Let $G=( S_1, S_2, i, T, p, \gamma)$ be a featured priority game
  structure and define $I: S\to( \Bool( N)\to M)$ by
  $I( s)( \phi)=( 0,\dotsc, 0)$ for all $s\in S$, $\phi\in \Bool( N)$.
  Let $p\in \px$, then player~1 wins the parity game on
  $\proj{ p}{ G}$ iff $\fpattr^*( I)( i)( \gamma_p)\ne \top$.
\end{theorem}

\section{Optimal Featured Strategies}
\label{se:opt}

A player-1 strategy in a game is said to be \emph{optimal} if it
realizes the value of the game against any player-2 strategy.  That
is, in a game with boolean objective such as reachability, energy, or
parity games, an optimal strategy for player~1 ensures that she wins
the game against any player-2 strategy \emph{if it is at all possible
  for her to win the game}.

In a game with quantitative objective, such as minimum reachability
games or discounted games, an optimal player-1 strategy
$\tilde \theta_1$ is one which realizes the value of the game against
any player-2 strategy, that is, such that the value
$\sup_{ \theta_2\in \Theta_2} \valr( \out{ \tilde \theta_1}{
  \theta_2})= \inf_{ \theta_1\in \Theta_1} \sup_{ \theta_2\in
  \Theta_2} \valr( \out{ \theta_1}{ \theta_2})$ for reachability
games; %
\linebreak 
$\inf_{ \theta_2\in \Theta_2} \vall{ \lambda}( \out{ \tilde \theta_1}{
  \theta_2})= \sup_{ \theta_1\in \Theta_1} \inf_{ \theta_2\in
  \Theta_2}$ $\vall{ \lambda}( \out{ \theta_1}{ \theta_2})$ for
discounted games.

We show how to compute optimal player-1 strategies for all featured
games introduced in the previous sections.

\subsection{Featured Reachability Games}

Let $G=( S_1, S_2, i, F, T)$ be a game structure.  A player-1 strategy
$\theta_1\in \Theta_1$ is \emph{memoryless} if it depends only on last
states of configurations, that is, if $\pend{ \pi}= \pend{ \pi'}$
implies $\theta_1( \pi)= \theta_1( \pi')$ for all
$\pi, \pi'\in \Conf_1$.  Hence memoryless player-1 strategies are
mappings $S_1\to S$.  It is well-known that it suffices to consider
memoryless strategies for reachability games.

Define again $I: S\to \Bool$ by $I( s)= \ltrue$ iff $s\in F$.  A
memoryless player-1 strategy $\theta_1: S_1\to S$ is \emph{locally
  optimal} if, for all $s\in S_1$,
$\attr^*( I)( s)= \attr^*( I)( \theta_1( s))$; that is, among all
options $s\to s'$, it $\theta_1( s)$ is such that $\attr^*( I)( s)=
\bigvee_{ s\to s'} \attr^*( I)( s')$ is maximized.

It is well-known that locally optimal strategies are optimal, hence if
player~1 wins the reachability game on $G$, then she can do so using a
locally optimal strategy.  Further, such a strategy can be trivially
extracted after the computation of $\attr^*$, hence optimal player-1
strategies in reachability games can be computed in polynomial time.

Now let $G=( S_1, S_2, i, F, T, \gamma)$ be a featured game structure.
We extend the domain of $\gamma: T\to \Bool( N)$ to finite paths in
$G$ by defining
$\gamma(( s_1,\dotsc, s_k))= \gamma((s_1, s_2))\land\dotsm\land
\gamma(( s_{ k- 1}, s_k))$.

A \emph{featured strategy} for player~$i$, for $i\in\{ 1, 2\}$, is a
function $\xi_i: \Conf_i\to( \Bool( N)\to S)$ such that for all
$\pi\in \Conf_i$ and $\phi\in \Bool( N)$,
$\pend{ \pi}\to \xi_i( \pi)( \phi)$.  The set of featured strategies
for player~$i$ is denoted $\Xi_i$.  We define mappings
$\Xi_i\times \Bool( N)\to \Theta_i$, denoted
$( \xi_i, \phi)\mapsto \xi_i( \phi)$ and defined by
$\xi_i( \phi)( \pi)= \xi_i( \pi)( \phi)$ for all $\pi\in \Conf_i$.

A pair of featured strategies $\xi_1\in \Xi_1$, $\xi_2\in \Xi_2$
defines a mapping $\out{ \xi_1}{ \xi_2}: \Bool( N)\to \ipaths{ G}$
from feature guards to infinite paths in $G$, where
$\out{ \xi_1}{ \xi_2}( \phi)=( s_1, s_2,\dotsc)$ is given by
\begin{equation*}
  s_1= i\,, \quad s_{ 2 k}= \xi_1( s_1,\dotsc, s_{ 2 k- 1})(
  \phi)\,, \quad s_{ 2 k+ 1}= \xi_2( s_1,\dotsc, s_{ 2 k})(
  \phi)\,.
\end{equation*}

Let $\phi\in \Bool( N)$.  Player~1 \emph{wins the $\phi$-reachability
  game} if there exists a strategy $\xi_1\in \Xi_1$ such that for all
$\xi_2\in \Xi_2$, with
$\out{ \xi_1}{ \xi_2}( \phi)=( s_1, s_2,\dotsc)$, there is an index
$k\ge 1$ for which $s_k\in F$ and
$\phi\land \gamma(( s_1,\dotsc, s_k))\nequiv \lfalse$.

\begin{lemma}
  \label{le:fval-val-reach-orig}
  Let $G=( S_1, S_2, i, F, T, \gamma)$ be a featured game structure
  and $p\in \px$.  Player~1 wins the reachability game in
  $\proj{ p}{ G}$ iff she wins the $\gamma_p$-reachability game in
  $G$.
\end{lemma}

A featured player-1 strategy $\xi_1\in \Xi_1$ is \emph{memoryless} if
$\pend{ \pi}= \pend{ \pi'}$ implies $\xi_1( \pi)= \xi_1( \pi')$ for
all $\pi, \pi'\in \Conf_1$.  Hence memoryless featured strategies are
mappings $S_1\to( \Bool( N)\to S)$.

Define $I: S\to( \Bool( N)\to \Bool)$ by $I( s)( \phi)= \ltrue$ if
$s\in F$; $\lfalse$ if $s\notin F$.  A memoryless featured player-1
strategy $\xi_1: S_1\to( \Bool( N)\to S)$ is \emph{locally optimal}
if, for all $s\in S_1$ and $\phi\in \Bool( N)$,
$\fattr^*( I)( s)( \phi)= \fattr^*( I)( \xi_1( s)( \phi))( \gamma(( s,
\xi_1( s)( \phi)))\land \phi)$.

\begin{theorem}
  \label{th:locoptfeat}
  Let $G$ be a featured game structure, then there exists a locally
  optimal player-1 strategy.  Further, if $\xi_1\in \Xi_1$ is locally
  optimal, then $\xi_1( \gamma_p)$ is optimal in $\proj{ p}{ G}$ for
  every $p\in \px$.
\end{theorem}

\subsection{Featured Minimum Reachability}

Let $G=( S_1, S_2, i, F, T)$ be a weighted game structure.  Memoryless
player-1 strategies are now mappings $\theta_1: S_1\to \Nat\times S$.
Such a strategy is locally optimal if
$\wattr^*( I)( s)= \theta_1( s)_1+ \wattr^*( I)( \theta_1( s)_2)$ for
all $s\in S_1$, where $I: S\to \Nat$ is defined by $I( s)= 0$ if
$s\in F$; $\infty$ if $s\notin F$, and
$\theta_1( s)=( \theta_1( s)_1, \theta_1( s)_2)$.

It is again well-known that locally optimal strategies are optimal,
hence optimal player-1 strategies in minimum reachability games can be
computed in pseudo-polynomial time.

Now let $G=( S_1, S_2, i, F, T, \gamma)$ be a featured weighted game
structure.  A featured player-$i$ strategy is a mapping
$\xi_i: \Conf_i\to( \Bool( N)\to \Nat\times S)$.  The outcome of a
pair $\xi_1$, $\xi_2$ of featured strategies is again a mapping
$\out{ \xi_1}{ \xi_2}: \Bool( N)\to \ipaths{ G}$ defined as expected.

The \emph{featured reachability value} of a mapping
$\pi: \Bool( N)\to \ipaths{ G}$ is the function
$\fvalr( \pi): \Bool( N)\to \Nat\cup\{ \infty\}$ given by
$\fvalr( \pi)( \phi)= \valr( \pi( \phi))$, and the \emph{featured
  minimum reachability value} of $G$ is
$\fvalr( G)= \inf_{ \xi_1\in \Xi_1} \sup_{ \xi_2\in \Xi_2} \fvalr(
\out{ \xi_1}{ \xi_2})$, where the order in
$\Bool( N)\to \Nat\cup\{ \infty\}$ is point-wise.

\begin{lemma}
  \label{le:fval-val-reach}
  For $G=( S_1, S_2, i, F, T, \gamma)$ any featured weighted game
  structure and $p\in \px$, $\valr( \proj{ p}{ G})= \fvalr( G)(
  \gamma_p)$.
\end{lemma}

Define $I: S\to( \Bool( N)\to \Nat\cup\{ \infty\})$ by
$I( s)( \phi)= 0$ if $s\in F$; $\lfalse$ if $s\notin F$.  Memoryless
featured player-1 strategies are mappings
$\xi_1: S_1\to( \Bool( N)\to \Nat\times S)$.  Such a strategy is
locally optimal if, for all $s\in S_1$ and $\phi\in \Bool( N)$,
$\fwattr^*( I)( s)( \phi)= \xi_1( s)( \phi)_1+ \fwattr^*( I)( \xi_1(
s)( \phi)_2)( \gamma(( s, \xi_1( s)( \phi)_1, \xi_1( s)(
\phi)_2))\land \phi)$.

\begin{theorem}
  Let $G$ be a featured weighted game structure, then there exists a
  locally optimal player-1 strategy.  Further, if $\xi_1\in \Xi_1$ is
  locally optimal, then $\xi_1( \gamma_p)$ is optimal in
  $\proj{ p}{ G}$ for every $p\in \px$.
\end{theorem}

\subsection{Featured Discounted Games}

Let $G=( S_1, S_2, i, T)$ be a weighted game structure,
$0< \lambda< 1$.  A memoryless player-1 strategy
$\theta_1: S_1\to \Int\times S$ is locally optimal if
$\dattr^*( s)= \theta_1( s)_1+ \lambda \dattr^*( \theta_1( s)_2)$ for
all $s\in S_1$.  Locally optimal strategies always exist and are
optimal~\cite{DBLP:journals/tcs/ZwickP96}.

Let $G=( S_1, S_2, i, T, \gamma)$ be a featured weighted game
structure.  The \emph{featured discounted value} of a mapping
$\pi: \Bool( N)\to \ipaths{ G}$ is the function
$\fvall{ \lambda}( \pi): \Bool( N)\to \Real$ given by
$\fvall{ \lambda}( \pi)( \phi)= \vall{ \lambda}( \pi( \phi))$.  The
featured discounted value of $G$ is
$\fvall{ \lambda}( G)= \sup_{ \xi_1\in \Xi_1}$
$\inf_{ \xi_2\in \Xi_2} \fvall{ \lambda}( \out{ \xi_1}{ \xi_2})$.

\begin{lemma}
  For $G=( S_1, S_2, i, T, \gamma)$ any featured weighted game
  structure and $p\in \px$,
  $\vall{ \lambda}( \proj{ p}{ G})= \fvall{ \lambda}( G)( \gamma_p)$.
\end{lemma}

A memoryless featured player-1 strategy
$\xi_1: S_1\to( \Bool( N)\to \Int\times S)$ is locally optimal if, for
all $s\in S_1$ and $\phi\in \Bool( N)$,
$\fdattr^*( s)( \phi)= \xi_1( s)( \phi)_1+ \lambda \fdattr^*( \xi_1(
s)( \phi)_2)( \gamma(( s, \xi_1( s)( \phi)_1, \xi_1( s)(
\phi)_2))\land \phi)$.

\begin{theorem}
  Let $G$ be a featured weighted game structure, then there exists a
  locally optimal player-1 strategy.  Further, if $\xi_1\in \Xi_1$ is
  locally optimal, then $\xi_1( \gamma_p)$ is optimal in
  $\proj{ p}{ G}$ for every $p\in \px$.
\end{theorem}

\subsection{Featured Energy Games}

Let $G=( S_1, S_2, i, T)$ be a weighted game structure and define
$M= \sum_{ s\in S} \max(\{ 0\}\cup\{ -x\mid( s, x, s')\in T\})$,
$W=\{ 0,\dotsc, M, \top\}$, and $I: S\to W$ by $I( s)= 0$ for all
$s\in S$ as before.  A memoryless player-1 strategy
$\theta_1: S_1\to \Int\times S$ is locally optimal if
$\eattr^*( I)( s)= \eattr^*( I)( \theta_1( s)_2)\ominus \theta_1(
s)_1$ for all $s\in S_1$.  If player~1 wins the energy game on $G$
with initial credit $v_0\in \Nat$, then she can do so using a locally
optimal strategy~\cite{DBLP:journals/fmsd/BrimCDGR11}.

Let $G=( S_1, S_2, i, T, \gamma)$ be a featured weighted game
structure, $v_0: \Bool( N)\to \Nat$, and $\phi\in \Bool( N)$.
Player~1 wins the $\phi$-energy game with initial credit $v_0$
if there exists a featured strategy $\xi_1\in \Xi_1$ such that for all
$\xi_2\in \Xi_2$, $\out{ \xi_1}{ \xi_2}( \phi)$ is energy positive
with initial credit $v_0( \phi)$.

\begin{lemma}
  Let $G=( S_1, S_2, i, T, \gamma)$ be a featured weighted game
  structure, $v_0: \Bool( N)\to \Nat$, and $p\in \px$.  Player~1 wins
  the energy game with initial credit $v_0( \gamma_p)$ in
  $\proj{ p}{ G}$ iff player~1 wins the $\gamma_p$-energy game in $G$
  with initial credit $v_0$.
\end{lemma}

Define $I: S\to( \Bool( N)\to W)$ by $I( s)( \phi)= 0$ for all
$s\in S$, $\phi\in \Bool( N)$.  A memoryless featured player-1
strategy $\xi_1: S_1\to( \Bool( N)\to \Int\times S)$ is locally
optimal if, for all $s\in S_1$ and $\phi\in \Bool( N)$,
$\feattr^*( I)( s)( \phi)= \feattr^*( I)( \xi_1( s)( \phi)_2)(
\gamma(( s, \xi_1( s)( \phi)_1, \xi_1( s)( \phi)_2))\land \phi)\ominus
\xi_1( s)( \phi)_1$.

\begin{theorem}
  Let $G$ be a featured weighted game structure, then there exists a
  locally optimal player-1 strategy.  Further, if $\xi_1\in \Xi_1$ is
  locally optimal, then $\xi_1( \gamma_p)$ is optimal in
  $\proj{ p}{ G}$ for every $p\in \px$.
\end{theorem}

\subsection{Featured Parity Games}

Let $G=( S_1, S_2, i, T, p)$ be a priority game structure,
$d= \max\{ p( s)\mid s\in S\}$ and $M\subseteq \Nat^d\cup\{ \top\}$ as
in Section~\ref{se:parity}, and define $I: S\to M$ by
$I( s)=( 0,\dotsc, 0)$ for all $s\in S$.  A memoryless player-1
strategy $\theta_1: S_1\to S$ is locally optimal if
$\pattr^*( I)( s)= \prog( \pattr^*( I), s, \theta_1( s))$ for all
$s\in S_1$.  If player~1 wins the parity game on $G$, then she can do
so using a locally optimal
strategy~\cite{DBLP:conf/stacs/Jurdzinski00}.

Let $G=( S_1, S_2, i, T, p, \gamma)$ be a featured priority game
structure and $\phi\in \Bool( N)$.  Player~1 wins the $\phi$-parity
game on $G$ if there exists a featured strategy $\xi_1\in \Xi_1$ such
that for all $\xi_2\in \Xi_2$, $\prio{ \out{ \xi_1}{ \xi_2}( \phi)}$
is an even number.

\begin{lemma}
  Let $G=( S_1, S_2, i, T, p, \gamma)$ be a featured priority game
  structure and $p\in \px$.  Player~1 wins the parity game in $\proj{
    p}{ G}$ iff player~1 wins the $\gamma_p$-parity game in $G$.
\end{lemma}

Define $I: S\to( \Bool( N)\to M)$ by $I( s)( \phi)=( 0,\dotsc, 0)$ for
all $s\in S$, $\phi\in \Bool( N)$.  A memoryless featured player-1
strategy $\xi_1: S_1\to( \Bool( N)\to S)$ is locally optimal if, for
all $s\in S_1$ and $\phi\in \Bool( N)$,
$\fpattr^*( I)( s)( \phi)= \fprog( \fpattr^*( I), s, \xi_1( s)(
\phi))( \gamma(( s, \xi_1( s))\land \phi)$.

\begin{theorem}
  Let $G$ be a featured weighted game structure, then there exists a
  locally optimal player-1 strategy.  Further, if $\xi_1\in \Xi_1$ is
  locally optimal, then $\xi_1( \gamma_p)$ is optimal in
  $\proj{ p}{ G}$ for every $p\in \px$.
\end{theorem}

\section{Symbolic Computation}
\label{se:comp}

The goal of feature-based analysis is to compute properties of an FTS
representation of a SPL for all products at once, and to do so in a
family-based way.  We have seen that for the various types of games we
have treated, values and optimal strategies may be computed by
calculating closures of attractors.  Hence we expose below
feature-based algorithms for calculating these closures.

\subsection{Featured Reachability Games}

Let $G=( S_1, S_2, i, F, T, \gamma)$ be a featured game structure and
define $I: S\to( \Bool( N)\to \Bool)$ by $I( s)( \phi)= \ltrue$ if
$s\in F$; $\lfalse$ if $s\notin F$.  Conceptually, the procedure for
calculating $J= \fattr^*( I)$ is a fixed-point algorithm: initialize
$J:= I$ and update $J:= J\vee \fattr( J)$ until $J$ stabilizes.

In order to symbolically represent functions from $\Bool( N)$, we use
guard partitions, see also~\cite{DBLP:journals/sttt/FahrenbergL19}.  A
\emph{guard partition} of $\px$ is a set $P\subseteq \Bool( N)$ such
that $\sem{ \bigvee P}= \px$, $\sem \phi\ne \emptyset$ for all
$\phi\in P$, and $\sem{ \phi_1}\cap \sem{ \phi_2}= \emptyset$ for all
$\phi_1, \phi_2\in P$ with $\phi_1\ne \phi_2$.  The set of all guard
partitions of $\px$ is denoted $\GPart\subseteq 2^{ \Bool( N)}$.

\begin{figure}[tbp]
  \begin{algorithmic}[1]
    \Function{Reduce}{$f: P\to X$}: $P'\to X$
    \State $P', f'\gets \emptyset$
    \While {$P\ne \emptyset$}
    \State Pick and remove $\phi$ from $P$
    \State $x\gets f( \phi)$
    \ForAll {$\psi\in P$}
    \If {$f( \psi)= x$}
    \State $\phi\gets \phi\lor \psi$
    \State $P\gets P\setminus\{ \psi\}$
    \EndIf
    \EndFor
    \State $P'\gets P'\cup\{ \phi\}$
    \State $f'( \phi)\gets x$
    \EndWhile
    \State \Return $f': P'\to X$
    \EndFunction
  \end{algorithmic}
  \Description{Algorithm which computes canonicalization.}
  \caption{Algorithm which computes canonicalization.}
  \label{fi:alg-reduce}
\end{figure}

\begin{figure}[tbp]
  \centering
  \begin{algorithmic}[1]
    \Function{Land}{$f_1: P_1\to \Bool, f_2: P_2\to \Bool$}: $P\to
    \Bool$
    \State $P, f\gets \emptyset$
    \ForAll {$\phi_1\in P_1$}
    \ForAll {$\phi_2\in P_2$}
    \If {$\sem{ \phi_1\land \phi_2}\ne\emptyset$}
    \State $P\gets P\cup\{ \phi_1\land \phi_2\}$
    \State $f( \gamma_1\land \gamma_2)\gets f_1( \gamma_1)\land f_2(
    \gamma_2)$
    \EndIf
    \EndFor
    \EndFor
    \State \Return \textsc{Reduce}($f$)
    \EndFunction
  \end{algorithmic}
  \Description{Algorithm for logical and.}
  \caption{Algorithm for logical and.}
  \label{fi:alg-land}
\end{figure}

\begin{figure}[tbp]
  \centering
  \begin{algorithmic}[1]
    \Function{Lor}{$f_1: P_1\to \Bool, f_2: P_2\to \Bool$}: $P\to
    \Bool$
    \State $P, f\gets \emptyset$
    \ForAll {$\phi_1\in P_1$}
    \ForAll {$\phi_2\in P_2$}
    \If {$\sem{ \phi_1\land \phi_2}\ne\emptyset$}
    \State $P\gets P\cup\{ \phi_1\land \phi_2\}$
    \State $f( \gamma_1\land \gamma_2)\gets f_1( \gamma_1)\lor f_2(
    \gamma_2)$
    \EndIf
    \EndFor
    \EndFor
    \State \Return \textsc{Reduce}($f$)
    \EndFunction
  \end{algorithmic}
  \Description{Algorithm for logical or.}
  \caption{Algorithm for logical or.}
  \label{fi:alg-lor}
\end{figure}

A function $f: P\to X$, for $P\in \GPart$ and $X$ any set, is
\emph{canonical} if $f( \phi_1)= f( \phi_2)$ implies $\phi_1= \phi_2$
for all $\phi_1, \phi_2\in P$.  A function $f: P\to X$ which is
\emph{not} canonical may be reduced into an equivalent canonical
function $f': P'\to X$ using the algorithm shown in
Fig.~\ref{fi:alg-reduce}.  Every function $\Bool( N)\to X$ has a
unique representation as a canonical function $P\to X$ for some $P\in
\GPart$.

The function for featured computation of attractors is shown in
Fig.~\ref{fi:alg-fattr}.  It uses the functions \textsc{Land} and
\textsc{Lor}, shown in Figs.~\ref{fi:alg-land} and~\ref{fi:alg-lor},
which compute logical operations on functions $P\to \Bool$: for
$f_1: P_1\to \Bool$ and $f_2: P_2\to \Bool$, \textsc{Land} returns
$f'= f_1\land f_2$, and \textsc{Lor} returns $f'= f_1\lor f_2$.

\begin{figure}[tbp]
  \begin{algorithmic}[1]
    \Function{Fattr}{$U: S\to( P\to \Bool)$}: $S\to( P'\to \Bool)$
    \State $U'\gets \emptyset$
    \ForAll {$s\in S$}
    \State $P_s', U_s'\gets \emptyset$
    \ForAll {$s\to s'$}
    \State {$Q_{ s'}, V_{ s'}\gets \emptyset$}
    \While {$P_{ s'}\ne \emptyset$}
    \State Pick and remove $\phi$ from $P_{ s'}$
    \State $\psi\gets \gamma(( s, s'))\land
    \phi$ \label{al:fattr.restrict}
    \If {$\sem{ \psi}\ne \emptyset$}
    \State $Q_{ s'}\gets Q_{ s'}\cup\{ \psi\}$
    \State $V_{ s'}( \psi)\gets U_{ s'}( \phi)$
    \EndIf
    \EndWhile
    \State $Q_{ s'}\gets Q_{ s'}\cup\{ \neg \gamma(( s, s'))\}$
    \State $V_{ s'}( \neg \gamma(( s, s')))\gets \lfalse$
    \If {$s\in S_1$} \label{al:fattr.combine}
    \State $U_s'\gets \textsc{Lor}( U_s', V_{ s'})$
    \EndIf
    \If {$s\in S_2$}
    \State $U_s'\gets \textsc{Land}( U_s', V_{ s'})$
    \EndIf
    \EndFor
    \State $U_s'\gets \textsc{Reduce}( U_s')$
    \EndFor
    \State \Return $U': S\to( P'\to \Bool)$
    \EndFunction
  \end{algorithmic}
  \Description{Computation of $\fattr$.}
  \caption{Computation of $\fattr$.}
  \label{fi:alg-fattr}
\end{figure}

The function \textsc{Fattr} in Fig.~\ref{fi:alg-fattr} computes one
iteration of $\fattr$ for all states $s\in S$.  It does so by
traversing all transitions $s\to s'$ (note that $s'$ might be equal to
$s$), restricting the partitions at $s'$ to $\gamma(( s, s'))$
(line~\ref{al:fattr.restrict}), and then computing
$U_s'= \bigvee_{ s\to s'} V_{ s'}$ or
$U_s'= \bigwedge_{ s\to s'} V_{ s'}$, depending on whether $s\in S_1$
or $s\in S_2$, in lines~\ref{al:fattr.combine}f.  The algorithm for
the fixed-point iteration to compute $\fattr^*$ is, then, shown in
Fig.~\ref{fi:alg-fattr*}.

\begin{figure}[tbp]
  \centering
  \begin{algorithmic}[1]
    \Function{Fattr*}{$G=( S_1, S_2, i, F, T, \gamma)$}: $S\to( P\to
    \Bool)$
    \State $J= \emptyset$
    \ForAll {$s\in S$}
    \State $U_s\gets\{ \ltrue\}$
    \If {$s\in F$}
    \State $J_s( \ltrue)\gets \ltrue$
    \Else
    \State $J_s( \ltrue)\gets \lfalse$
    \EndIf
    \EndFor
    \Repeat
    \State $J_\text{old}\gets J$
    \State $J\gets \textsc{Lor}( J, \textsc{Fattr}( J))$
    \Until {$J= J_\text{old}$}
    \State \Return $J$
    \EndFunction
  \end{algorithmic}
  \Description{Fixed-point iteration for $\fattr^*$.}
  \caption{Fixed-point iteration for $\fattr^*$.}
  \label{fi:alg-fattr*}
\end{figure}

\subsection{Featured Minimum Reachability Games}

Let $G=( S_1, S_2, i, F, T, \gamma)$ be a featured weighted game
structure define $I: S\to( \Bool( N)\to \Nat\cup\{ \infty\})$ by
$I( s)( \phi)= 0$ if $s\in F$; $\infty$ if $s\notin F$.  The
computation of the fixed point $\fwattr^*( I)$ is similar to the one
in the previous section and shown in Figs.~\ref{fi:alg-min}
through~\ref{fi:alg-fwattr*}.  In the algorithm for $\fwattr$
(Fig.~\ref{fi:alg-fwattr}), line~\ref{al:fwattr.iter} now adds the
weights of the respective transitions, and the logical operations have
been replaced by maximum and minimum (Figs.~\ref{fi:alg-min}
and~\ref{fi:alg-max}).

\begin{figure}[tbp]
  \centering
  \begin{algorithmic}[1]
    \Function{Min}{$f_1: P_1\to \Nat\cup\{ \infty\}, f_2: P_2\to
      \Nat\cup\{ \infty\}$}: $P\to \Nat\cup\{ \infty\}$
    \State $P, f\gets \emptyset$
    \ForAll {$\phi_1\in P_1$}
    \ForAll {$\phi_2\in P_2$}
    \If {$\sem{ \phi_1\land \phi_2}\ne\emptyset$}
    \State $P\gets P\cup\{ \phi_1\land \phi_2\}$
    \State $f( \gamma_1\land \gamma_2)\gets \min( f_1( \gamma_1),
    f_2( \gamma_2))$
    \EndIf
    \EndFor
    \EndFor
    \State \Return \textsc{Reduce}($f$)
    \EndFunction
  \end{algorithmic}
  \Description{Algorithm for minimum.}
  \caption{Algorithm for minimum.}
  \label{fi:alg-min}
\end{figure}

\begin{figure}[tbp]
  \centering
  \begin{algorithmic}[1]
    \Function{Max}{$f_1: P_1\to \Nat\cup\{ \infty\}, f_2: P_2\to
      \Nat\cup\{ \infty\}$}: $P\to \Nat\cup\{ \infty\}$
    \State $P, f\gets \emptyset$
    \ForAll {$\phi_1\in P_1$}
    \ForAll {$\phi_2\in P_2$}
    \If {$\sem{ \phi_1\land \phi_2}\ne\emptyset$}
    \State $P\gets P\cup\{ \phi_1\land \phi_2\}$
    \State $f( \gamma_1\land \gamma_2)\gets \max( f_1( \gamma_1),
    f_2( \gamma_2))$
    \EndIf
    \EndFor
    \EndFor
    \State \Return \textsc{Reduce}($f$)
    \EndFunction
  \end{algorithmic}
  \Description{Algorithm for maximum.}
  \caption{Algorithm for maximum.}
  \label{fi:alg-max}
\end{figure}

\begin{figure}[tbp]
  \begin{algorithmic}[1]
    \Function{Fwattr}{$U: S\to( P\to \Nat\cup\{ \infty\})$}: $S\to(
    P'\to \Nat\cup\{ \infty\})$
    \State $U'\gets \emptyset$
    \ForAll {$s\in S$}
    \State $P_s', U_s'\gets \emptyset$
    \ForAll {$s\tto{ x} s'$}
    \State {$Q_{ s'}, V_{ s'}\gets \emptyset$}
    \While {$P_{ s'}\ne \emptyset$}
    \State Pick and remove $\phi$ from $P_{ s'}$
    \State $\psi\gets \gamma(( s, s'))\land \phi$
    \If {$\sem{ \psi}\ne \emptyset$}
    \State $Q_{ s'}\gets Q_{ s'}\cup\{ \psi\}$
    \State $V_{ s'}( \psi)\gets x+ U_{ s'}(
    \phi)$ \label{al:fwattr.iter}
    \EndIf
    \EndWhile
    \State $Q_{ s'}\gets Q_{ s'}\cup\{ \neg \gamma(( s, s'))\}$
    \State $V_{ s'}( \neg \gamma(( s, s')))\gets \lfalse$
    \If {$s\in S_1$} \label{al:fwattr.combine}
    \State $U_s'\gets \textsc{Min}( U_s', V_{ s'})$
    \EndIf
    \If {$s\in S_2$}
    \State $U_s'\gets \textsc{Max}( U_s', V_{ s'})$
    \EndIf
    \EndFor
    \State $U_s'\gets \textsc{Reduce}( U_s')$
    \EndFor
    \State \Return $U': S\to( P'\to \Nat\cup\{ \infty\})$
    \EndFunction
  \end{algorithmic}
  \Description{Computation of $\fwattr$.}
  \caption{Computation of $\fwattr$.}
  \label{fi:alg-fwattr}
\end{figure}

\begin{figure}[tbp]
  \centering
  \begin{algorithmic}[1]
    \Function{Fwattr*}{$G=( S_1, S_2, i, F, T, \gamma)$}: $S\to(
    P\to \Nat\cup\{ \infty\})$
    \State $J= \emptyset$
    \ForAll {$s\in S$}
    \State $U_s\gets\{ \ltrue\}$
    \If {$s\in F$}
    \State $J_s( \ltrue)\gets 0$
    \Else
    \State $J_s( \ltrue)\gets \infty$
    \EndIf
    \EndFor
    \Repeat
    \State $J_\text{old}\gets J$
    \State $J\gets \textsc{Min}( J, \textsc{Fwattr}( J))$
    \Until {$J= J_\text{old}$}
    \State \Return $J$
    \EndFunction
  \end{algorithmic}
  \Description{Fixed-point iteration for $\fwattr^*$.}
  \caption{Fixed-point iteration for $\fwattr^*$.}
  \label{fi:alg-fwattr*}
\end{figure}

\subsection{Featured Discounted Games}

The algorithms for computing values of featured discounted games are
shown in Figs.~\ref{fi:alg-fdattr} and~\ref{fi:alg-fdattr*}.  They use
functions \textsc{Min} and \textsc{Max} similar to the ones in
Figs.~\ref{fi:alg-min} and~\ref{fi:alg-max}.  The function
\textsc{Fdattr*} in Fig.~\ref{fi:alg-fdattr*} takes a discounting
factor $\lambda$ and a precision $\epsilon$ as inputs; $\lambda$ is
used for the iteration in \textsc{Fdattr}, and $\epsilon$ is used to
terminate the computation of $\fdattr^*$ once a desired level of
precision has been reached.

\begin{figure}[tbp]
  \begin{algorithmic}[1]
    \Function{Fdattr}{$U: S\to( P\to \Real), \lambda$}: $S\to( P'\to
    \Real)$
    \State $U'\gets \emptyset$
    \ForAll {$s\in S_1$}
    \State $P_s', U_s'\gets \emptyset$
    \ForAll {$s\tto{ x} s'$}
    \State {$Q_{ s'}, V_{ s'}\gets \emptyset$}
    \While {$P_{ s'}\ne \emptyset$}
    \State Pick and remove $\phi$ from $P_{ s'}$
    \State $\psi\gets \gamma(( s, s'))\land \phi$
    \If {$\sem{ \psi}\ne \emptyset$}
    \State $Q_{ s'}\gets Q_{ s'}\cup\{ \psi\}$
    \State $V_{ s'}( \psi)\gets x+ \lambda U_{ s'}(
    \phi)$ \label{al:fdattr.iter}
    \EndIf
    \EndWhile
    \State $Q_{ s'}\gets Q_{ s'}\cup\{ \neg \gamma(( s, s'))\}$
    \State $V_{ s'}( \neg \gamma(( s, s')))\gets \lfalse$
    \If {$s\in S_1$} \label{al:fdattr.combine}
    \State $U_s'\gets \textsc{Max}( U_s', V_{ s'})$
    \EndIf
    \If {$s\in S_2$}
    \State $U_s'\gets \textsc{Min}( U_s', V_{ s'})$
    \EndIf
    \EndFor
    \State $U_s'\gets \textsc{Reduce}( U_s')$
    \EndFor
    \State \Return $U': S\to( P'\to \Real)$
    \EndFunction
  \end{algorithmic}
  \Description{Computation of $\fdattr$.}
  \caption{Computation of $\fdattr$.}
  \label{fi:alg-fdattr}
\end{figure}

\begin{figure}[tbp]
  \centering
  \begin{algorithmic}[1]
    \Function{Fdattr*}{$G=( S_1, S_2, i, T, \gamma), \lambda, \epsilon$}:
    $S\to( P\to \Real)$
    \State $J= \emptyset$
    \ForAll {$s\in S$}
    \State $U_s\gets\{ \ltrue\}$
    \State $J_s( \ltrue)\gets 0$
    \EndFor
    \Repeat
    \State $J_\text{old}\gets J$
    \State $J\gets \textsc{Fdattr}( J, \lambda)$
    \Until {$\| J- J_\text{old}\|< \epsilon$}
    \State \Return $J$
    \EndFunction
  \end{algorithmic}
  \Description{Fixed-point iteration for $\fdattr^*$.}
  \caption{Fixed-point iteration for $\fdattr^*$.}
  \label{fi:alg-fdattr*}
\end{figure}

\subsection{Featured Energy and Parity Games}

The algorithms for computing the attractors of featured energy games
and of featured parity games are very similar to the ones already
shown and not depicted due to space restrictions.  They can be found
in the long version~\cite{journals/corr/FahrenbergL20}.





\section{Conclusion}

We have in this work lifted most of the two-player games which are
used in model checking and controller synthesis to software product
lines.  We have introduced featured versions of reachability games,
minimum reachability games, discounted games, energy games, and parity
games.  We have shown how to compute featured attractors for these
games, using family-based algorithms with late splitting, and how to
use these featured attractors to compute winners, values, and optimal
strategies for all products at once.

The astute reader may have noticed that \emph{mean-payoff games} are
conspicuously absent from this paper.  The immediate reason for this
absence is that mean-payoff games do not admit attractors; instead
they are solved by computing loops~\cite{DBLP:journals/tcs/ZwickP96}.
\cite{DBLP:journals/fmsd/BrimCDGR11}~show an easy reduction from
mean-payoff to energy games which may be used to compute winners in
featured mean-payoff games.  To compute values and optimal strategies,
the reduction to discounted games
in~\cite{DBLP:conf/birthday/GimbertZ08}, building on earlier work
in~\cite{DBLP:journals/tcs/ZwickP96}, may be used.

Two-player games are an established technique for model checking and
control synthesis, and our work shows that this technology may be
lifted to featured model checking and featured control synthesis.  In
future work we plan to implement our algorithms and integrate them
into the mCRL2 toolset~\cite{DBLP:conf/tacas/BunteGKLNVWWW19,
  DBLP:conf/fase/BeekVW17}, using BDD representations of product
families, in order to evaluate our work on benchmark models.

We also plan to extend our work into the probabilistic and timed
settings.  Controller synthesis often deals with real-time or hybrid
systems, and SPL models of such systems are by now
well-established~\cite{DBLP:conf/splc/BeekLLV15,
  DBLP:conf/hase/RodriguesANLCSS15, DBLP:conf/splc/CordySHL12}.  For
real-time systems, we are looking into extending \emph{timed
  games}~\cite{DBLP:conf/cav/BehrmannCDFLL07} with features,
analogously to the featured timed automata
of~\cite{DBLP:conf/splc/CordySHL12}; for probabilistic systems, a
featured extension of stochastic games~\cite{books/PetersV87} appears
straight-forward.

\bibliographystyle{plain}
\bibliography{mybib}

\iflongversion
\newpage

\section*{Appendix: Proofs}

\begin{proof}[Proof of Thm.~\ref{th:fattr}]
  Let $H= \proj{ p}{ G}=( S_1, S_2, i, F, T')$ and, for clarity, write
  $\attr_H$ for the attractor in $H$ and $\fattr_G$ for the one in
  $G$.  We need to show that
  $\fattr_G^*( I)( i)( \gamma_p)= \attr_H^*( I)( i)$.  (Note that we
  are using the same notation $I$ for both $G$ and $H$; this should
  cause no confusion.)

  The conclusion will follow once we can show that for all $n\ge 0$
  and all $s\in S$,
  $\fattr_G^n( I)( s)( \gamma_p)= \attr_H^n( I)( s)$.  We do so by
  induction on $n$.  For $n= 0$ both sides of the equation become
  $\ltrue$ iff $s\in F$, so this is clear.

  Now let $n\ge 0$ and assume that for all $s'\in S$,
  $\fattr_G^n( I)( s')( \gamma_p)= \attr_H^n( I)( s')$.  Let
  $s\in S_1$, then
  \begin{align*}
    \fattr_G^{ n+ 1}( I)( s)( \gamma_p) &= \textstyle\bigvee_{ s\ttto{
        x}{ G} s'} \fattr_G^n( I)( s')( \gamma(( s, x, s'))\land
    \gamma_p) \\
    &= \textstyle\bigvee_{ s\ttto{ x}{ H} s'} \fattr_G^n( I)( s')(
    \gamma_p) \\
    &= \textstyle\bigvee_{ s\ttto{ x}{ H} s'} \attr_H^n( I)( s')=
    \attr_H^{ n+ 1}( I)( s)\,;
  \end{align*}
  for $s\in S_2$ the proof is similar.
\end{proof}

\begin{proof}[Proof of Thm.~\ref{th:fwattr}]
  Let $H= \proj{ p}{ G}=( S_1, S_2, i, F, T')$; we need to prove that
  $\fwattr_G^*( I)( i)( \gamma_p)= \wattr_H^*( I)( i)$.  We show
  inductively that for all $n\ge 0$ and all $s\in S$,
  $\fwattr_G^n( I)( s)( \gamma_p)= \wattr_H^n( I)( s)$, which will
  imply the conclusion.  For $n= 0$ both sides of the equation become
  $0$ if $s\in F$ and $\infty$ otherwise, so the base case is clear.

  Now let $n\ge 0$ and assume that for all $s'\in S$,
  $\fwattr_G^n( I)( s')( \gamma_p)= \wattr_H^n( I)( s')$.  Let
  $s\in S_1$, then
  $\fwattr_G^{ n+ 1}( I)( s)( \gamma_p)= \min_{ s\ttto{ x}{ G} s'} x+
  \fwattr_G^n( I)( s')( \gamma(( s, x, s'))\land \gamma_p)= \min_{
    s\ttto{ x}{ H} s'} x+ \fwattr_G^n( I)( s')( \gamma_p)= \min_{
    s\ttto{ x}{ H} s'} x+ \wattr_H^n( I)( s')= \wattr_H^{ n+ 1}( I)(
  s)$; for $s\in S_2$ the proof is similar.
\end{proof}

\begin{proof}[Proof of Thm.~\ref{th:fdattr}]
  Define a metric on $S\to( \Bool( N)\to \Real)$ by
  $d( U_1, U_2)= \max_{ s\in S}$
  $\max_{ \phi\in \Bool( N)}| U_1( s)( \phi)- U_2( s)( \phi)|$.  Then\linebreak
  $d( \fdattr( U_1), \fdattr( U_2))\le \lambda d( U_1, U_2)$ for any
  two functions $U_1, U_2$, that is, $\fdattr$ is a \emph{contraction}
  on the complete metric space $S_1\to \Real^{ \Bool( N)}$.  By the
  Banach fixed-point theorem, $\fdattr$ has a unique fixed point which
  is $\fdattr^*$.

  Let $p\in \px$ and $H= \proj{ p}{ G}=( S_1, S_2, i, T')$; we need to
  show that $\fattr^*_H( i)= \fdattr^*_G( i)( \gamma_p)$.  Now for any
  $U: S\to( \Bool( N)\to \Real)$ and $s\in S_1$,
  $\fdattr( U)( s)( \gamma_p)= \max_{ s\ttto{ x}{ G} s'} x+ \lambda U(
  s')( \gamma(( s, x, s'))\land \gamma_p)= \max_{ s\ttto{ x}{ H} s'}
  x+ \lambda U( s')( \gamma_p)$, and the same can be shown if
  $s\in S_2$ instead.  Hence the equation systems defining
  $\dattr^*_H$ and $\fdattr^*_G( \cdot)( \gamma_p)$ are the same;
  consequently, also their unique fixed points are equal.
\end{proof}

\begin{proof}[Proof of Thm.~\ref{th:feattr}]
  Let $H= \proj{ p}{ G}=( S_1, S_2, i, T')$; we need to prove that
  $\feattr_G^*( I)( i)( \gamma_p)= \eattr_H^*( I)( i)$.  We show
  inductively that for all $n\ge 0$ and all $s\in S$,
  $\feattr_G^n( I)( s)( \gamma_p)= \eattr_H^n( I)( s)$, which will
  imply the conclusion.  For $n= 0$, the equation becomes
  $I( s)( \gamma_p)= I( s)$ which is clear.

  Now let $n\ge 0$ and assume that for all $s'\in S$,
  $\feattr_G^n( I)( s')( \gamma_p)= \eattr_H^n( I)( s')$.  Let
  $s\in S_1$, then
  $\feattr_G^{ n+ 1}( I)( s)( \gamma_p)= \min_{ s\ttto{ x}{ G} s'}$\linebreak
  $\feattr_G^n( I)( s')( \gamma(( s, x, s'))\land \gamma_p)\ominus x=
  \min_{ s\ttto{ x}{ H} s'} \feattr_G^n( I)( s')( \gamma_p)\ominus x=
  \min_{ s\ttto{ x}{ H} s'} \eattr_H^n( I)( s')\ominus x= \eattr_H^{
    n+ 1}( I)( s)$; similarly for $s\in S_2$.
\end{proof}

\begin{proof}[Proof of Thm.~\ref{th:fpattr}]
  Let $H= \proj{ p}{ G}=( S_1, S_2, i, T', p)$; we need to prove that
  $\fpattr_G^*( I)( i)( \gamma_p)= \pattr_H^*( I)( i)$.  We show
  inductively that for all $n\ge 0$ and all $s\in S$,
  $\fpattr_G^n( I)( s)( \gamma_p)= \pattr_H^n( I)( s)$, which will
  imply the conclusion.  For $n= 0$, the equation becomes
  $I( s)( \gamma_p)= I( s)$ which is clear.

  Now let $n\ge 0$ and assume that for all $s'\in S$,
  $\fpattr_G^n( I)( s')( \gamma_p)= \pattr_H^n( I)( s')$.  Let
  $s\in S_1$, then $\fpattr_G^{ n+ 1}( I)( s)( \gamma_p)=$ %
  \linebreak 
  $\max\{ \fpattr_G^n( I)( s)( \gamma_p),$
  $\min_{ s\ttto{}{ G} s'} \fprog( \fpattr_G^n( I), s, s')( \gamma((
  s, s'))\land \gamma_p)\}= \max\{ \pattr_H^n( I)( s),$
  $\min_{ s\ttto{}{ H} s'} \fprog( \fpattr_G^n( I), s, s')(
  \gamma_p)\}= \max\{ \pattr_H^n( I)( s), \min_{ s\ttto{}{ H} s'}$
  $\min\{ m\in M\mid m\succeq_{ p( s)}$ %
  \linebreak 
  $\fpattr_G^n( I)( s')(
  \gamma_p)\}\}= \max\{ \pattr_H^n( I)( s),$
  $\min_{ s\ttto{}{ H} s'} \min\{ m\in M\mid m\succeq_{ p( s)}
  \pattr_H^n( I)( s')\}\}= \max\{ \pattr_H^n( I)( s),$
  $\min_{ s\ttto{}{ H} s'}$ %
  \linebreak 
  $\prog( \pattr_H^n( I), s, s')\}= \pattr_H^{
    n+ 1}( I)( s)$.  For $s\in S_2$ the reasoning is similar.
\end{proof}

\begin{proof}[Proof of Lemma~\ref{le:fval-val-reach-orig}]
  Assume that player~1 wins the reachability game in
  $H= \proj{ p}{ G}$.  Then there is $\theta_1\in \Theta_1$ such that
  for all $\theta_2\in \Theta_2$, writing
  $\out{ \theta_1}{ \theta_2}=( s_1, s_2,\dotsc)$, there is an index
  $k\ge 1$ for which $s_k\in F$.  Let $\theta_2\in \Theta_2$.  All
  transitions $( s_1, s_2),( s_2, s_3),\dotsc$ are in $H$, hence
  $p\models \gamma(( s_1,\dotsc, s_k))$,
  \ie~$\gamma_p\land \gamma(( s_1,\dotsc, s_k))\nequiv \lfalse$.  Let
  $\xi_1\in \Xi_1$ be any strategy for which
  $\xi_1( \gamma_p)= \theta_1$.  We have shown that for any
  $\xi_2\in \Xi_2$, writing
  $\out{ \xi_1}{ \xi_2}( \phi)=( s_1, s_2,\dotsc)$, there is an index
  $k\ge 1$ for which $s_k\in F$ and
  $\phi\land \gamma(( s_1,\dotsc, s_k))\nequiv \lfalse$; that is,
  player~1 wins the $\gamma_p$-reachability game in $G$.

  For the converse, assume that player~1 wins the
  $\gamma_p$-reachability game in $G$, and let $\xi_1\in \Xi_1$ be
  such that for any $\xi_2\in \Xi_2$, writing
  $\out{ \xi_1}{ \xi_2}( \gamma_p)=( s_1, s_2,\dotsc)$, there is an
  index $k\ge 1$ for which $s_k\in F$ and
  $\gamma_p\land \gamma(( s_1,\dotsc, s_k))\nequiv \lfalse$,
  \ie~$p\models \gamma(( s_1,\dotsc, s_k))$.  Then
  $p\models \gamma(( s_1, s_2))\land\dotsm\land \gamma(( s_{ k- 1},
  s_k))$, so that all the transitions
  $( s_1, s_2),\dotsm,( s_{ k- 1}, s_k)$ are present in $H$.  Let
  $\theta_1= \xi_1( \gamma_p)$.  We have shown that for all
  $\theta_2\in \Theta_2$, writing
  $\out{ \theta_1}{ \theta_2}=( s_1, s_2,\dotsc)$, there is an index
  $k\ge 1$ for which $s_k\in F$; that is, player~1 wins the
  reachability game in $H$.
\end{proof}

\begin{proof}[Proof of Thm.~\ref{th:locoptfeat}]
  We show the second claim first.  Write $H= \proj{ p}{ G}$, assume
  $\xi_1$ to be locally optimal, write $\theta_1= \xi_1( \gamma_p)$,
  and let $s\in S_1$.  Then
  $\attr_H^*( I)( s)= \fattr_G^*( I)( s)( \gamma_p)= \fattr_G^*( I)(
  \theta_1( s))$ $( \gamma(( s, \theta_1( s)))\land \gamma_p)=
  \fattr_G^*( I)( \theta_1( s))( \gamma_p)= \attr_H^*( I)( \theta_1(
  s))$, thus $\theta_1$ is locally optimal in $\proj{ p}{ G}$.

  For the first claim of the theorem, let $s\in S_1$ and
  $\phi\in \Bool( N)$.  We have
  $\fattr^*( I)( s)( \phi)= \bigvee_{ s\to s'} \fattr^*( I)( s')(
  \gamma(( s, s'))\land \phi)$.  The set $\{ s'\in S_2\mid s\to s'\}$
  is finite, hence there is $\tilde s'$ such that
  $\fattr^*( I)( s)( \phi)= \fattr^*( I)( \tilde s')( \gamma(( s,
  \tilde s'))\land \phi)$.  Define $\xi_1( s)( \phi)= \tilde
  s'$.
\end{proof}

\begin{proof}[Proof of Lemma~\ref{le:fval-val-reach}]
  Denoting strategy sets in $G$ by $\Xi_i$ and in $\proj{ p}{ G}$ by
  $\Theta_i$, we see that $\Theta_i= \Xi_i( \gamma_p)$.  Then
  $\fvalr( G)( \gamma_p)= \inf_{ \xi_1\in \Xi_1} \sup_{ \xi_2\in
    \Xi_2}$
  $\fvalr( \out{ \xi_1}{ \xi_2})( \gamma_p)= \inf_{ \xi_1\in \Xi_1}
  \sup_{ \xi_2\in \Xi_2}$ %
  \linebreak 
  $\valr( \out{ \xi_1}{ \xi_2}( \gamma_p))=
  \inf_{ \xi_1\in \Xi_1} \sup_{ \xi_2\in \Xi_2} \valr( \out{ \xi_1(
    \gamma_p)}{ \xi_2( \gamma_p)})= \inf_{ \theta_1\in \Xi_1(
    \gamma_p)} \sup_{ \theta_2\in \Xi_2( \gamma_p)}$
  $\valr( \out{ \theta_1}{ \theta_2})= \inf_{ \theta_1\in \Theta_1}
  \sup_{ \theta_2\in \Theta_2}$ %
  \linebreak 
  $\valr( \out{ \theta_1}{ \theta_2})=
  \valr( \proj{ p}{ G})$.
\end{proof}

\newpage

\section*{Appendix: Other Algorithms}

\begin{figure}[bp]
  \begin{algorithmic}[1]
    \Function{Feattr}{$U: S\to( P\to W)$}: $S\to( P'\to W)$
    \State $U'\gets \emptyset$
    \ForAll {$s\in S_1$}
    \State $P_s', U_s'\gets \emptyset$
    \ForAll {$s\tto{ x} s'$}
    \State {$Q_{ s'}, V_{ s'}\gets \emptyset$}
    \While {$P_{ s'}\ne \emptyset$}
    \State Pick and remove $\phi$ from $P_{ s'}$
    \State $\psi\gets \gamma(( s, s'))\land \phi$
    \If {$\sem{ \psi}\ne \emptyset$}
    \State $Q_{ s'}\gets Q_{ s'}\cup\{ \psi\}$
    \State $V_{ s'}( \psi)\gets U_{ s'}( \phi)\ominus x$
    \label{al:feattr.iter}
    \EndIf
    \EndWhile
    \State $Q_{ s'}\gets Q_{ s'}\cup\{ \neg \gamma(( s, s'))\}$
    \State $V_{ s'}( \neg \gamma(( s, s')))\gets \lfalse$
    \If {$s\in S_1$} \label{al:feattr.combine}
    \State $U_s'\gets \textsc{Min}( U_s', V_{ s'})$
    \EndIf
    \If {$s\in S_2$}
    \State $U_s'\gets \textsc{Max}( U_s', V_{ s'})$
    \EndIf
    \EndFor
    \State $U_s'\gets \textsc{Reduce}( U_s')$
    \EndFor
    \State \Return $U': S\to( P'\to W)$
    \EndFunction
  \end{algorithmic}
  \Description{Computation of $\feattr$.}
  \caption{Computation of $\feattr$.}
  \label{fi:alg-feattr}
\end{figure}

\begin{figure}[tbp]
  \centering
  \begin{algorithmic}[1]
    \Function{Feattr*}{$G=( S_1, S_2, i, T, \gamma)$}: $S\to(
    P\to \Nat\cup\{ \infty\})$
    \State $J= \emptyset$
    \ForAll {$s\in S$}
    \State $U_s\gets\{ \ltrue\}$
    \State $J_s( \ltrue)\gets 0$
    \EndFor
    \Repeat
    \State $J_\text{old}\gets J$
    \State $J\gets \textsc{Max}( J, \textsc{Feattr}( J))$
    \Until {$J= J_\text{old}$}
    \State \Return $J$
    \EndFunction
  \end{algorithmic}
  \Description{Fixed-point iteration for $\feattr^*$.}
  \caption{Fixed-point iteration for $\feattr^*$.}
  \label{fi:alg-feattr*}
\end{figure}

\begin{figure}[tbp]
  \begin{algorithmic}[1]
    \Function{Fprog}{$U: S\to( P\to M), s, s'$}: $P'\to M$
    \State $U', P'\gets \emptyset$
    \While {$P_{ s'}\ne \emptyset$}
    \State Pick and remove $\phi$ from $P_{ s'}$
    \State $P'\gets P'\cup\{ \phi\}$
    \State $U'( \phi)\gets \min\{ m\in M\mid m\succeq_{ p( s)} U_{
      s'}( \phi)\}$
    \EndWhile
    \State \Return \textsc{Reduce}($U'$)
    \EndFunction
  \end{algorithmic}
  \Description{Computation of $\fprog$.}
  \caption{Computation of $\fprog$.}
  \label{fi:alg-fprog}
\end{figure}

\begin{figure}[tbp]
  \begin{algorithmic}[1]
    \Function{Fpattr}{$U: S\to( P\to M)$}: $S\to( P'\to M)$
    \State $U'\gets \emptyset$
    \ForAll {$s\in S$}
    \State $P_s', U_s'\gets \emptyset$
    \ForAll {$s\to s'$}
    \State {$Q_{ s'}, V_{ s'}\gets \emptyset$}
    \While {$P_{ s'}\ne \emptyset$}
    \State Pick and remove $\phi$ from $P_{ s'}$
    \State $\psi\gets \gamma(( s, s'))\land \phi$
    \If {$\sem{ \psi}\ne \emptyset$}
    \State $Q_{ s'}\gets Q_{ s'}\cup\{ \psi\}$
    \State $V_{ s'}( \psi)\gets \textsc{Fprog}( U, s, s')( \phi)$
    \label{al:fpattr.iter}
    \EndIf
    \EndWhile
    \State $Q_{ s'}\gets Q_{ s'}\cup\{ \neg \gamma(( s, s'))\}$
    \State $V_{ s'}( \neg \gamma(( s, s')))\gets \lfalse$
    \If {$s\in S_1$} \label{al:fpattr.combine}
    \State $U_s'\gets \textsc{Min}( U_s', V_{ s'})$
    \EndIf
    \If {$s\in S_2$}
    \State $U_s'\gets \textsc{Max}( U_s', V_{ s'})$
    \EndIf
    \EndFor
    \State $U_s'\gets \textsc{Reduce}( U_s')$
    \EndFor
    \State \Return $U': S\to( P'\to M)$
    \EndFunction
  \end{algorithmic}
  \Description{Computation of $\fpattr$.}
  \caption{Computation of $\fpattr$.}
  \label{fi:alg-fpattr}
\end{figure}

\begin{figure}[tbp]
  \centering
  \begin{algorithmic}[1]
    \Function{Fpattr*}{$G=( S_1, S_2, i, T, p, \gamma)$}: $S\to( P\to
    M)$
    \State $J= \emptyset$
    \ForAll {$s\in S$}
    \State $U_s\gets\{ \ltrue\}$
    \State $J_s( \ltrue)\gets 0$
    \EndFor
    \Repeat
    \State $J_\text{old}\gets J$
    \State $J\gets \textsc{Max}( J, \textsc{Fpattr}( J))$
    \Until {$J= J_\text{old}$}
    \State \Return $J$
    \EndFunction
  \end{algorithmic}
  \Description{Fixed-point iteration for $\fpattr^*$.}
  \caption{Fixed-point iteration for $\fpattr^*$.}
  \label{fi:alg-fpattr*}
\end{figure}

\else
\fi

\end{document}